\documentclass[fleqn,usenatbib]{mnras}

\usepackage{newtxtext,newtxmath}
\usepackage{xcolor, soul}
\definecolor{teagreen}{rgb}{0.82, 0.94, 0.75}
\sethlcolor{teagreen}

\usepackage[T1]{fontenc}

\DeclareRobustCommand{\VAN}[3]{#2}
\let\VANthebibliography\thebibliography
\def\thebibliography{\DeclareRobustCommand{\VAN}[3]{##3}\VANthebibliography}

\usepackage{graphicx}	
\usepackage{amsmath}	

\usepackage{stackengine}

\title[WISDOM project -- {\MakeUppercase{\romannumeral 11}}. SFE in NGC~3169]{WISDOM project -- {\MakeUppercase{\romannumeral 11}}. Star Formation Efficiency in the Bulge of the AGN-host Galaxy NGC~3169 with SITELLE and ALMA}

\author[A.\ Lu et al.]{
Anan Lu,$^{1}$\thanks{E-mail: anan.lu@mail.mcgill.ca}
Hope Boyce,$^{1}$
Daryl Haggard,$^{1}$
Martin Bureau,$^{2,}$ $^{3}$
Fu-Heng Liang,$^{2}$
Lijie Liu,$^{4}$
Woorak Choi,$^{3}$
\newauthor
Michele Cappellari,$^{2}$
Laurent Chemin,$^{5}$
M{\'e}lanie Chevance,$^{6}$
Timothy A. Davis,$^{7}$
Laurent Drissen,$^{8}$
\newauthor
Jacob S. Elford,$^{7}$
Jindra Gensior,$^{9}$
J.~M.~Diederik Kruijssen,$^{6}$
Thomas Martin,$^{8}$
Etienne Mass{\'e},$^{8}$
\newauthor
Carmelle Robert,$^{8}$
Ilaria Ruffa,$^{7}$
Laurie Rousseau-Nepton,$^{10}$
Marc Sarzi,$^{11}$
Gabriel Savard,$^{8}$
\newauthor
Thomas G. Williams$^{12}$
\\
$^{1}$McGill Space Institute and Department of Physics, McGill University, 3600 rue University, Montreal, QC H3A 2T8, Canada\\
$^{2}$Sub-department of Astrophysics, Department of Physics, University of Oxford, Denys Wilkinson Building, Keble Road, Oxford OX1 3RH, UK\\
$^{3}$Yonsei Frontier Lab and Department of Astronomy, Yonsei University, 50 Yonsei-ro, Seodaemun-gu, Seoul 03722, Republic of Korea\\
$^{4}$DTU-Space, Technical University of Denmark, Elektrovej 327, DK-2800 Kgs. Lyngby, Denmark\\
$^{5}$Centro de Astronom\'ia - CITEVA, Universidad de Antofagasta, Avenida Angamos 601, Antofagasta, 1270300, Chile\\
$^{6}$Astronomisches Rechen-Institut, Zentrum f{\"u}r Astronomie der Universit{\"a}t Heidelberg, M{\"o}nchhofstraße 12-14, 69120 Heidelberg, Germany\\
$^{7}$Cardiff Hub for Astrophysics Research \& Technology, School of Physics \& Astronomy, Cardiff University, Queens Buildings, The Parade, Cardiff, CF24 3AA,UK\\
$^{8}$D{\'e}partement de Physique, de G{\'e}nie Physique et d’Optique, Universit{\'e} Laval, Qu{\'e}bec, QC G1V 0A6, Canada\\
$^{9}$Institute for Computational Science, University of Z{\"u}rich, Winterthurerstrasse 190, 8057 Z{\"u}rich, Switzerland\\
$^{10}$Canada–France–Hawaii Telescope, Kamuela, HI 96743, USA\\
$^{11}$Armagh Observatory and Planetarium, College Hill, Armagh BT61 9DG, UK\\
$^{12}$Max Planck Institut f{\"u}r Astronomie, K{\"o}nigstuhl 17, 69117 Heidelberg, Germany\\
}

\date{Accepted XXX. Received YYY; in original form ZZZ}

\pubyear{2021}

\begin{document}
\label{firstpage}
\maketitle

\begin{abstract}
  The star formation efficiency (SFE) has been shown to vary across different environments, particularly within galactic starbursts and deep within the bulges of galaxies. Various quenching mechanisms may be responsible, ranging from galactic dynamics to feedback from active galactic nuclei (AGN). Here, we use spatially-resolved observations of warm ionised gas emission lines (H$\beta$, [\ion{O}{iii}] $\lambda\lambda$4959,5007, [\ion{N}{ii}] $\lambda\lambda$6548,6583, H$\alpha$ and [\ion{S}{ii}] $\lambda\lambda$6716,6731) from the imaging Fourier transform spectrograph SITELLE at the Canada-France-Hawaii Telescope (CFHT) and cold molecular gas (\textsuperscript{12}CO(2-1)) from the Atacama Large Millimeter/sub-millimeter Array (ALMA) to study the SFE in the bulge of the AGN-host galaxy NGC~3169. After distinguishing star-forming regions from AGN-ionised regions using emission-line ratio diagnostics, we measure spatially-resolved molecular gas depletion times ($\tau_{\rm dep}\equiv$1/SFE) with a spatial resolution of $\approx100$~pc within a galactocentric radius of $1.8$~kpc. We identify a star-forming ring located at radii $1.25\pm0.6$~kpc with an average $\tau_{\rm dep}$ of 0.3 Gyr. At radii $<0.9$~kpc, however, the molecular gas surface densities and depletion times increase with decreasing radius, the latter reaching approximately $2.3$~Gyr at a radius $\approx500$~pc. Based on analyses of the gas kinematics and comparisons with simulations, we identify AGN feedback, bulge morphology and dynamics as the possible causes of the radial profile of SFE observed in the central region of NGC~3169.
\end{abstract}



\section{Introduction}
\label{intro}

The tight correlation between the star-formation rates (SFRs) and stellar masses ($M_\star$) of galaxies, referred to as the star-formation main sequence (SFMS; e.g.\ \citealt{brinchmann2004physical, noeske2007star, rodighiero2011lesser, speagle2014highly}), suggests a self-regulated evolution of galaxy mass and size \citep{bouche2010impact,genzel2010study,dave2012analytic,feldmann2015argo}. However, many galaxies deviate from the SFMS, particularly the "quenched" galaxy population, which has much lower specific star-formation rates (sSFRs; i.e.\ ${\rm SFR}/M_\star$). Our understanding of this star formation truncation remains poor, with galactic dynamics \citep[e.g.][]{martig2009morphological,martig2013atlas3d,gensior2020heart}, mergers \citep[e.g.][]{moustakas2013primus} and/or active galactic nuclei (AGN) feedback \citep[e.g.][]{pontzen2017quench} all offering possible contributing mechanisms. 

Fundamentally, stars should form from the cold and dense (typically molecular) gas in the interstellar medium (ISM). Star-formation quenching is therefore either the result of a low molecular gas fraction ($f_{\rm H_2}$; $f_{\rm H_2}\equiv\frac{M_{\rm H_2}}{M_\star}$) and/or a low star formation efficiency (SFE; ${\rm SFE}\equiv{\rm SFR}/M_{\rm H_2}$), as the reduction of either of them would lead to a lower sSFR. In turn, the depletion time ($\tau_{\rm dep}$; $\tau_{\rm dep}\equiv{\rm SFE}^{-1}=\frac{M_{\rm H_2}}{\rm SFR}$), i.e., the timescale required to exhaust the molecular gas present at the current SFR, is a useful quantity to probe star formation. The power-law relation between $M_{\rm H_2}$ and SFR on galactic scales \citep{kennicutt1998global,kennicutt2021revisiting} shows that $\tau_{\rm dep}$ of starburst galaxies is $10-10^2$~Myr, while those of spiral galaxies tend to be much longer, reaching over $10^3$~Myr. On sub-galactic scales, where $\tau_{\rm dep}$ is typically calculated as the ratio of the molecular gas mass surface density ($\Sigma_{\rm H_2}$) to the SFR surface density ($\Sigma_{\rm SFR}$), \citet{bigiel2008star} revealed a depletion time of $\approx2$~Gyr.
 
Within individual galaxies, evidence also suggests that the SFE is not uniform across the variety of local environments (e.g.\ disc, spheroid, spiral arms, inter-arm regions; \citealt{querejeta2019dense}). Contrary to the high SFEs of starburst regions, the SFEs are reduced in galaxy bulges \citep[e.g.][]{davis2014atlas3d,kruijssen2014controls}. The SFRs of early-type galaxies (ETGs) and bulges are also suppressed by a factor of $\approx 2.5$ compared to those of spiral galaxy discs with similar molecular gas mass surface densities \citep[e.g.][]{saintonge2011cold}. It is possible that this star formation suppression is analogous to that observed in the central regions of spirals and growing spheroids at higher redshifts \citep{lang2014bulge}, and it may therefore be a symptom of a more general process. Indeed, the central molecular zone (CMZ), i.e, the inner $\approx500$~pc in radius of our own Milky Way, also has a SFR one order of magnitude smaller than the Galactic average, despite very high molecular gas mass surface densities \citep[e.g.][]{longmore2013variations}. A multitude of physical processes have been implicated, but the high turbulent pressure within the CMZ is likely the most important \citep{kruijssen2014controls}.
 
It is therefore crucial to compare different environments to validate commonly-postulated quenching mechanisms, such as ``morphological quenching'', where a nearly spherical potential holds gas stable against star-formation \citep[e.g.][]{martig2009morphological,martig2013atlas3d,gensior2020heart}, dynamical processes such as shear and tides, that stabilise the gas clouds against gravitational collapse \citep[e.g.][]{davis2014atlas3d,2019MNRAS.484.5734K,liu2021wisdom}, and AGN feedback, whereby radiation and jets from AGN physically remove and/or heat up the gas \citep[e.g.][]{pontzen2017quench}.

To distinguish these different mechanisms in the inner regions of galaxies, we have initiated a project to measure the SFEs of galaxy bulges in a spatially-resolved manner, combining high-spatial resolution \textsuperscript{12}CO(2-1) measurements (by a cold gas indicator) from the Atacama Large Millimeter/sub-millimeter Array (ALMA) with ionised hydrogen emission-line measurements (by a SFR indicator) from the imaging Fourier transform spectrograph SITELLE at the Canada-France-Hawaii Telescope (CFHT). Here we present a pilot study based on a single galaxy, the nearby early type spiral galaxy NGC~3169 with a mean SFR of $\approx2$~M$_\odot$~yr$^{-1}$. This galaxy has a dense bulge (with an effective radius $\approx1.4$~kpc and a S{\'e}rsic index of $\approx4.17$; from \citealt{dong2006low}) and a low-luminosity AGN (LLAGN; e.g.\ \citealt{ghosh2008low,mathur2008finding}), making it ideal for a deep investigation of various quenching mechanisms. The target and observations are described in Section~\ref{obs}, followed by surface-density and line-ratio maps in Section~\ref{ini_ana}. In Section~\ref{analysis}, we calculate depletion times and examine the radial trend of SFE and its correlation with galaxy dynamics (i.e.\ the virial parameters of the \ion{H}{ii} regions and galaxy velocity field). In Section~\ref{discussion}, we discuss the mechanisms contributing to the radial profile of SFE in the central region, including AGN feedback, bulge morphology and dynamics. We conclude and discuss future work in Section~\ref{conclusion}.

\begin{table}
  \caption{Key parameters of NGC~3169.}
  \label{tab:summary}
  \begin{tabular}{|c|l|l|}
    
    \hline
    Ref. & Quantity & Value \\
    \hline
    (1) & Type & SA(s)a pec \\
    \hline
    (2) & RA & $10^{\rm h}14^{\rm m}15\fs05$ \\
    \hline
    (3) & DEC & $3^\circ27\arcmin57\farcs90$\\
    \hline
    (4) & Distance (Mpc) & $18.7$\\
    \hline
    (5) & $\log({\rm M_{H_2}~/~M_\odot})$ & $9.32$\\
    \hline
    (6) & $\log({\rm \Sigma_{H_2,~2~kpc}~/~M_\odot~pc^{-2}})$ & $2.31$\\
    \hline
    (7) & $\log({\rm M_\star~/~M_\odot})$ & $10.84$ \\
    \hline
    (8) & $\sigma_\star$ (km~s$^{-1}$) & $163$\\
    \hline
    (9) & $\log({\rm \mu_\star~/~M_\odot~kpc^{-2}})$ & $8.26$ \\
    \hline
    (10) & $\log({\rm sSFR~/~yr^{-1}})$  & $-10.55$ \\
    \hline
    (11) & $R_{\rm e,KS}$ (\arcsec) & $85.7$\\
    \hline
    (12) & $\log({\rm SFR~/~M_\odot~yr^{-1}})$ & $0.29$\\
    \hline
    (13) & $\log({\rm \Sigma_{SFR,~2~kpc}~/~M_\odot~yr^{-1}~kpc^{-2}})$ & $-0.67$\\
    \hline
    (14) & $L_{\rm Edd}$ (erg~s$^{-1}$) & $\approx10^{46}$ \\
    \hline
  \end{tabular}
  
  {\it Notes:} {(1) Morphology of the galaxy taken from \citet{1991rc3..book.....D}. (2) and (3) are the RA-DEC coordinate of the galaxy center used for SITELLE observations and referenced for all the maps in this work. (4) Assumed distance taken from the NASA extragalactic database redshift independent distance catalogue \citep{steer2016redshift}. (5) Molecular gas mass measured within our ALMA field of view as described in Section~\ref{gas} (6) Mean molecular gas surface density measured within the inner $2$~kpc of the galaxy, as described in Section~\ref{gas}. (7) Stellar mass of the galaxy taken from \citet{leroy2019z}. (8) Stellar velocity dispersion taken from \citet{heraudeau1999stellar}. (9) Stellar mass surface density ($\mu_\star={\rm M_\star}/(2\pi R_e^2)$) estimated within the effective radius ($R_{\rm e,KS}$) of the galaxy. (10) Specific star formation rate (sSFR=${\rm M_\star/SFR}$) calculated based on (7) and (12). (11) $K_s$-band effective radius estimated from the $2~\mu$m all-sky survey (2MASS; \citealt{jarrett20032mass}). (12) Total star formation rate taken from \citet{leroy2019z}. (13) Mean star formation rate surface density measured within the inner $2$~kpc of the galaxy, as described in Section~\ref{SFR}. (14) Estimated Eddington luminosity of the supermassive black hole, taken from \citet{ghosh2008low}.}
  
\end{table}


\section{Observations and data reduction}
\label{obs}


\subsection{NGC~3169}
\label{target}

\begin{figure*}
  \centering
  \includegraphics[width=0.85\textwidth]{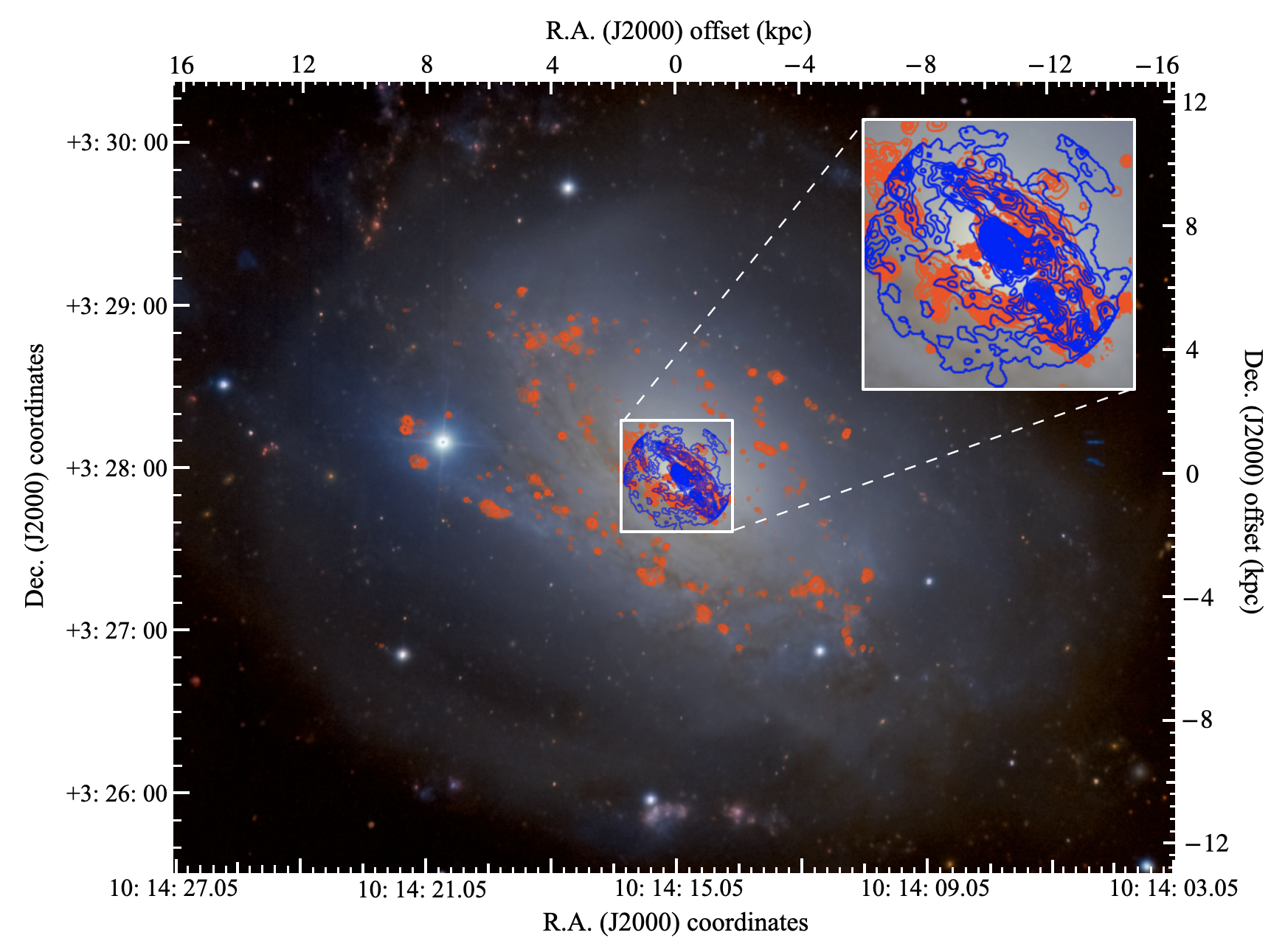}
  \caption{\label{fig:HST} Deep image of NGC~3169 extracted from the CFHT SITELLE observations discussed in Section~\ref{Halpha}, with the H$\alpha$ integrated intensity overlaid as orange contours and the CO integrated intensity (extracted from the ALMA observations discussed in Section~\ref{CO}) overlaid as blue contours. The white box indicates the central $\approx3.6\times3.6$~kpc$^2$ region, encompassing the galaxy bulge, which is the main focus of this paper. The inset shows a zoom-in view of the central region.}
\end{figure*}

NGC~3169 is an S(a) spiral galaxy at a distance $D=18.7$~Mpc, according to the NASA Extragalactic Database redshift-independent distance catalogue \citep{steer2016redshift}. It has an inclination $i=54^\circ$, a position angle $PA=45^\circ$ and a systemic heliocentric velocity $V_{\rm sys}=1232$~km~s$^{-1}$ (from the \ion{H}{i} $21$-cm observations of \citealt{lee2012pre}, consistent with the best-fitting systemic velocity derived from our own CO data discussed in Section~\ref{vfield}). NGC~3169 is representative of early-type spiral galaxies, while interacting with its close companion galaxy NGC~3166 \citep{sil2006central}. It hosts a complex central region, consisting of a dense and clear bulge, a small-scale bar, an AGN and a young nucleus, making it possible to probe multiple components that can affect local SFE.
Figure~\ref{fig:HST} shows an deep image produced from our SITELLE observations of NGC~3169, overlaid with contours of H$\alpha$ flux (measured with SITELLE) and \textsuperscript{12}CO(2-1) flux (measured with ALMA).

The \ion{H}{i} gas mass of NGC~3169 is $(4.2\pm0.5)\times10^{9}$~M$_\odot$ \citep{lee2012pre}. The ultraviolet and infrared images from the {\it Wide-field Infrared Survey Explorer} ({\it WISE}) and {\it Galaxy Evolution Explorer} ({\it GALEX}) missions \citep{leroy2019z} indicate a stellar mass of $\log (M_\star/{\rm M_\odot})\approx10.84$ and a star formation rate $\log(\rm SFR/{\rm M_\odot~yr^{-1}})\approx0.29$ within a central aperture of diameter $\approx19\arcmin$. Key parameters of NGC~3169 are summarised in Table~\ref{tab:summary}.

Probing the centre of NGC~3169, \citet{sil2006central} obtained spatially-resolved measurements in the wavelength range $4200$--$5600$~\AA~from the Multi Pupil Fiber Spectrograph (MPFS) at the prime focus of the 6-m Special Astrophysical Observatory (SAO) telescope. They reported a young stellar population with an estimated age of $1$--$2$~Gyr in the central region, including a star-like nucleus and a ring of young stars $6\arcsec$ ($\approx600$~pc) from the centre. Within the nuclear region (radius~$\lesssim200$~pc), \citet{dong2006low} reported an H$\alpha$ luminosity $\approx10^{39}$~ergs~s$^{-1}$, based on a spectroscopic survey carried out with the 5-m telescope at Mount Palomar by \citet{ho1997search}. Integrated emission-line ratios in this region were reported as H$\alpha$/H$\beta=5.03$, [\ion{O}{iii}]$\lambda$5007/H$\beta=2.88$ and [\ion{N}{ii}]$\lambda$6583/H$\alpha=2.07$, indicating the presence of an AGN. 

Furthermore, \citet{dong2006low} used the {\tt GALFIT} software \citep{peng2002detailed} on Two Micron All Sky Survey (2MASS) $K_{\rm s}$-band images to decompose the galaxy into a spheroidal bulge and an exponential disc. They reported bulge and disc $K_{\rm s}$-band integrated apparent magnitudes $m_{\rm bul}=7.52$ and $m_{\rm disc}=8.42$. The effective radius of the bulge component is $15\farcs7$ or $1.4$~kpc at our assumed distance. Based on this bulge-disc decomposition, we estimate the enclosed bulge-to-disk luminosity ratio (B/D) as a function of galactocentric radius; B/D approaches unity at a radius of $\approx660$~pc. This suggests that the bulge is dominating over the disc within $\approx660$~pc radius, while the impact from the bulge can reach over $1.4$~kpc.

The presence of a low-luminosity AGN (LLAGN) at the heart of NGC~3169 has been confirmed at multiple wavelengths. Based on \textit{Chandra} X-ray data, \citet{ghosh2008low} reported a hard X-ray detection and fitted the spectrum with an absorbed power law, deriving a power-law index $\Gamma\approx2$ and a hydrogen column density $N_{\rm H}\approx10^{23}$~cm$^{-2}$. They inferred an unabsorbed broad-band X-ray flux $f_{\rm 0.3-8~keV}=1\times10^{-11}$~erg~cm$^{-2}$~s$^{-1}$, equivalent to a luminosity $L_{\rm 0.3-8~keV}=1.1\times10^{41}$~erg~s$^{-1}$ at our assumed distance. A milliarcsecond-scale radio source was also detected, with a $5$-GHz flux $f_{\rm 5~GHz}=6.6\times10^{-26}$~erg~cm$^{-2}$~s$^{-1}$~Hz$^{-1}$. The measured brightness temperature exceeds $10^{7.7}$~K, ruling out a starburst and/or supernova remnant origin, thus confirming the presence of a LLAGN \citep{nagar2005radio}. Following the bulge-disc decomposition summarised above, \citet{dong2006low} estimated a central supermassive black hole mass $\log(M_\bullet/\rm M_\odot)=8.2$ using their derived bulge mass - black hole mass correlation based on 117 galaxies. Separately, \citet{heraudeau1999stellar} reported a central stellar velocity dispersion $\sigma_\ast=163$~km~s$^{-1}$, corresponding to $\log(M_\bullet/\rm M_\odot)=7.8$ using the stellar velocity dispersion - black hole mass correlation of \citet{kormendy2013coevolution}. Taking the average of these two black hole mass estimates, the Eddington luminosity is estimated to be $L_{\rm Edd}\approx10^{46}$~erg~s$^{-1}$ \citep{ghosh2008low}. With this well-characterised LLAGN at the centre, the potential influence of AGN feedback on star formation in the bulge of NGC~3169 can be uniquely probed.
 

\subsection{Ionised gas data}
\label{Halpha}

NGC~3169 was observed with SITELLE \citep{drissen2019sitelle} at CFHT during the 2020B semester (programme number 20Bc25). The observations were centred at R.A.~(J2000)$=10^{\rm h}14^{\rm m}15\fs05$ and Dec.~(J2000)$=3^\circ27\arcmin57\farcs90$. SITELLE is an optical imaging Fourier transform spectrograph (IFTS) equipped with two E2V detectors each with $2048 \times 2064$ pixels. The field of view (FOV) is $11\arcmin\times11\arcmin$, resulting in a mean spaxel size on the sky of $0\farcs31\times0\farcs31$ (or $28\times28$~pc$^2$). Two datacubes were obtained: one centred on the emission lines of [\ion{N}{ii}]$\lambda6548$, H$\alpha$, [\ion{N}{ii}]$\lambda6583$, [\ion{S}{ii}]$\lambda6716$ and [\ion{S}{ii}]$\lambda6731$ with the SN3 filter ($6480$--$6860$~\AA) at a mean spectral resolution of $R\approx2000$; the other centred on the emission lines of H$\beta$, [\ion{O}{iii}]$\lambda4959$ and [\ion{O}{iii}]$\lambda5007$ with the SN2 filter ($4840$--$5120$~\AA) at a mean spectral resolution $R\approx900$.

\begin{figure*}
\centering
\includegraphics[width=0.95\textwidth]{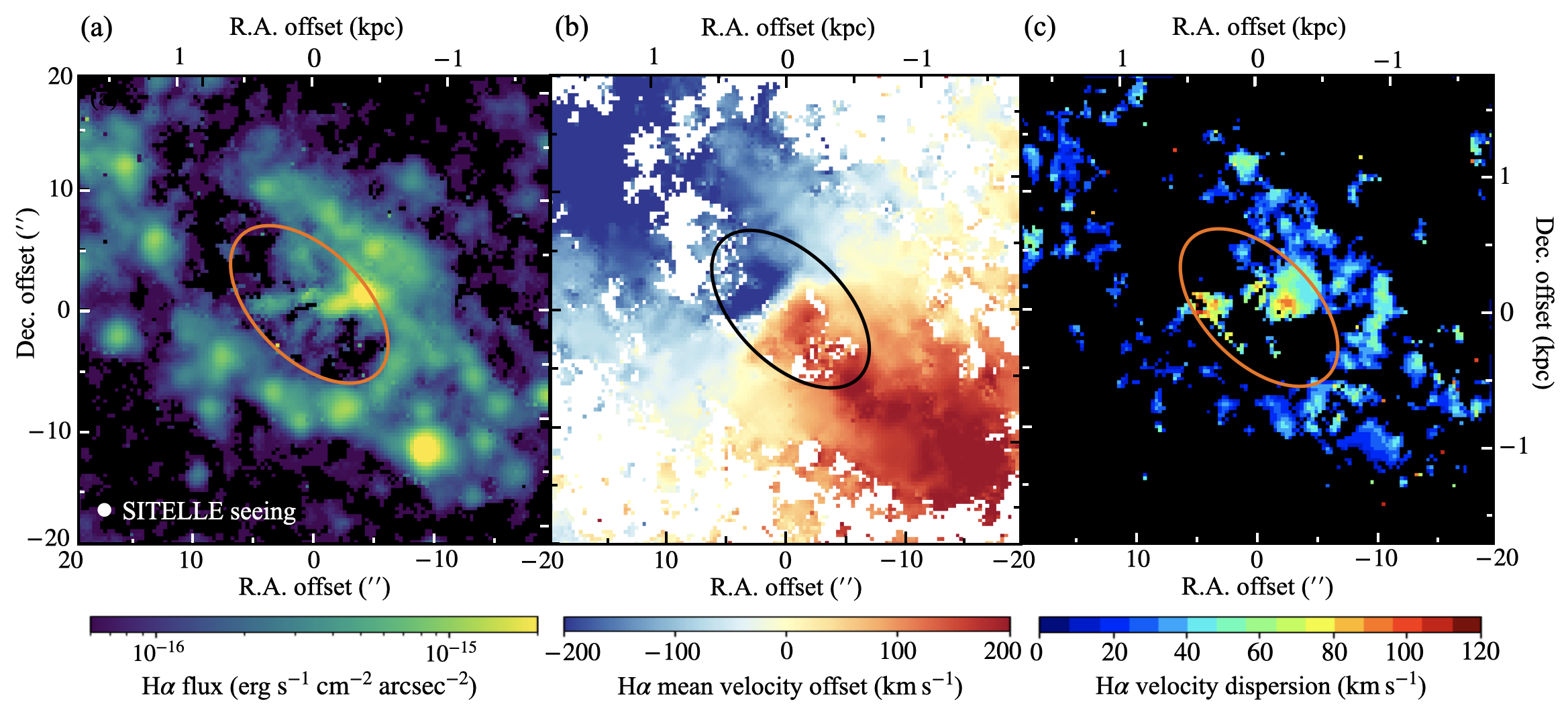}
\includegraphics[width=0.87\textwidth]{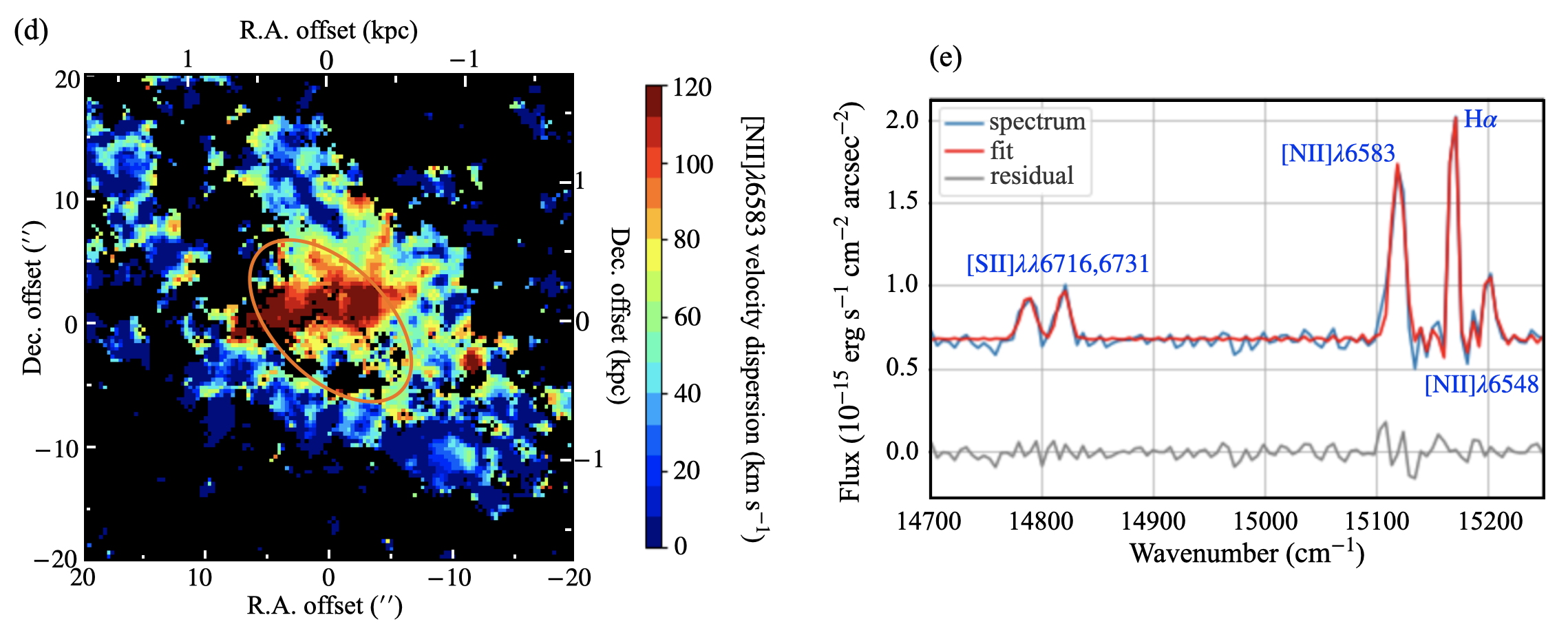}
\caption{Top row: H$\alpha$ maps of the central region of NGC~3169 ($\approx3.6\times3.6$~kpc$^2$), encompassing the galaxy bulge, extracted from our SITELLE SN3 datacube using the sinc-gauss line-fitting function in ORCS as described in Section~\ref{Halpha}. (a) Integrated intensity map. (b) Mean velocity map, with respect to the heliocentric velocity of the galaxy ($1232$~km~s$^{-1}$, see Section~\ref{target}). (c) Velocity dispersion map. Bottom row: (d) [\ion{N}{ii}]$\lambda6583$ velocity dispersion map of the same region. The orange and black ellipses in panels a, b, c and d separate the inner disc and outer star-forming ring discussed in Sections~\ref{analysis}--\ref{discussion}. (e) Sample spectrum (blue line) containing the emission lines [\ion{N}{ii}]$\lambda6548$, H$\alpha$, [\ion{N}{ii}]$\lambda6583$, [\ion{S}{ii}]$\lambda6716$ and [\ion{S}{ii}]$\lambda6731$. The ORCS-fitted spectrum is shown as a red line, while the residuals are shown as a grey line and shifted lower for illustration purposes.}
\label{fig:Halpha_data}
\end{figure*}

The data reduction was performed with the 
{\tt ORBS} software developed for SITELLE \citep{martin2015orbs,martin2021data}. The seeing was measured to be $1\farcs1$ (or $99$~pc) from the full-width at half-maxima (FWHM) of Gaussian fits to foreground stars from the Gaia catalogue \citep{lindegren2018gaia}. The data were further calibrated in wavelength based on velocity measurements of the OH sky line. As a final pre-processing step, sky background subtraction was performed using a median sky spectrum extracted from a $200\times200$ spaxels region located far away from the galaxy disc. The emission lines in each spaxel were then fitted using the extraction software {\tt ORCS} \citep{martin2015orbs}. For each emission line, maps of flux, mean velocity and velocity dispersion were generated from {\tt ORCS} output parameters including integrated flux within each spaxel, maximum intensity, velocity, FWHM, continuum level and signal-to-noise ratio ($S/N$) of the spectrum as well as their uncertainties. Modelling the sinc-Gauss profile of the emission lines allows to measure velocity dispersions much smaller than the nominal spectral resolution, provided that the $S/N$ is sufficient \citep{martin2016optimal}. A detection threshold was then applied based on the $3\sigma$ noise level of the H$\alpha$+[\ion{N}{ii}] flux. For illustration purposes, the full set of H$\alpha$ maps (before extinction correction) and the [\ion{N}{ii}]$\lambda6583$ velocity dispersion map used for this work are presented in Figure~\ref{fig:Halpha_data}. The extinction correction and conversion of the H$\alpha$ flux to $\Sigma_{\rm SFR}$ are discussed in Section~\ref{SFR}. We use the ratios of these emission lines to correctly identify the \ion{H}{ii} regions and examine the extent of the AGN impact in Sections~\ref{line_ratio} and \ref{discussion}.

\subsection{Molecular gas data}
\label{CO}

The NGC~3169 molecular gas data used were obtained in the $^{12}$CO(2-1) line (230 GHz) using ALMA in January 2015 (configuration: C36-4; programme: 2015.1.00598.S; PI: Bureau) and January 2016 (configuration: 7M; programme: 2016.2.00053.S; PI: Liu), as part of the mm-Wave Interferometric Survey of Dark Object Masses (WISDOM) project \citep{onishi2017wisdom}.
The first observation (2015.1.00598.S) obtained $605$~s on the source using thirty-eight $12$-m antennae, resulting in a primary beam (FWHM) of $27\farcs4$ and a maximum recoverable scale of $5\farcs8$. The intrinsic channel width was $1.27$~km~s$^{-1}$ ($977$~kHz). The line was covered by a correlator window of $1.875$~GHz, equivalent to $2450$~km~s$^{-1}$. The second observation obtained $1120$~s on source using eight $7$-m antennae, that extended the primary beam (FWHM) to $45\farcs7$ and the maximum recoverable scale to $36\farcs4$. The line was covered by a correlator window of $2$~GHz.

The raw ALMA data were reduced using the Common Astronomy Software Applications ({\tt CASA}) pipeline (version 5.3), which automatically processes the data by performing basic flagging and calibration \citep{mcmullin2007casa}. The two ALMA observations were then combined using the {\tt CASA} task {\tt concat}. Continuum emission was detected, measured over the full line-free bandwidth, and then subtracted from the data in the $uv$ plane using the {\tt CASA} task {\tt uvcontsub}. The data were then imaged using the {\tt CASA} task {\tt tclean} (version 5.6.1), adopting a channel width of $2$~km~s$^{-1}$, a Briggs' weighting robust parameter of $2.0$ and a $uv$-taper of $0\farcs7$, thus achieving a synthesised beam size (FWHM) of $0\farcs99\times0\farcs84$ ($89$~pc$\times76$~pc) at a position angle of $-58^\circ$, roughly matching the seeing of the H$\alpha$ observations. Finally, we convolved this datacube spatially by a narrow and slightly elongated two-dimensional Gaussian, to achieve a perfectly circular synthesized beam of FWHM $1\farcs1$, thus perfectly matching the seeing of the H$\alpha$ observations. This produced our final fully calibrated and continuum-subtracted $^{12}$CO(2-1) datacube of NGC~3169, with a mean RMS noise level of $3.06$~mJy~beam$^{-1}$ per $2$~km~s$^{-1}$ channel, although we also performed primary beam correction for the imaging of moment maps. Sky coordinates and spaxel size of the datacube were also matched to those of the SITELLE datacubes. Finally, another datacube was produced following the same procedure but with a channel width of $15$~km~s$^{-1}$, to enhance the S/N for the purpose of the gas kinematics analyses carried out in Section~\ref{vfield}.

\begin{figure*}
\centering
\includegraphics[width=0.95\textwidth]{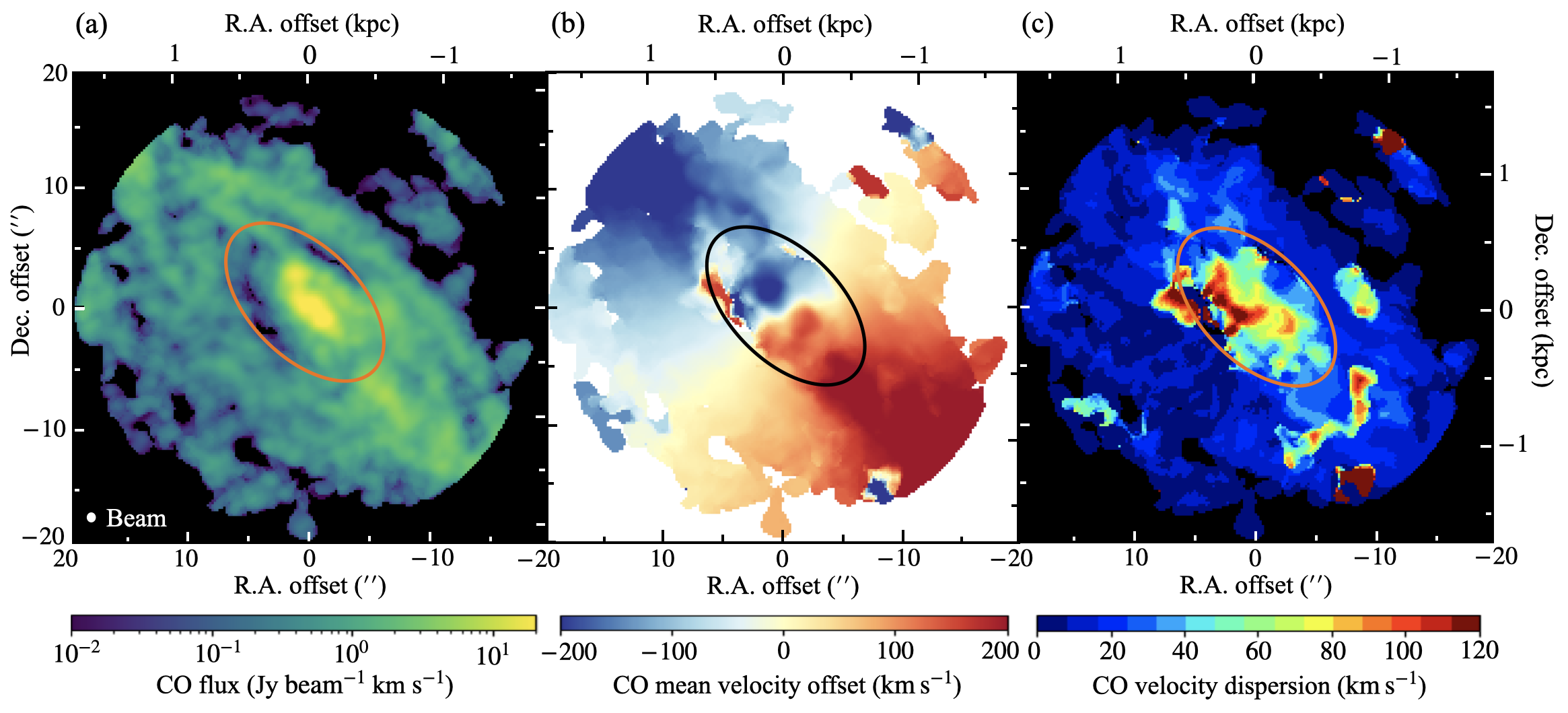}
\includegraphics[width=0.87\textwidth]{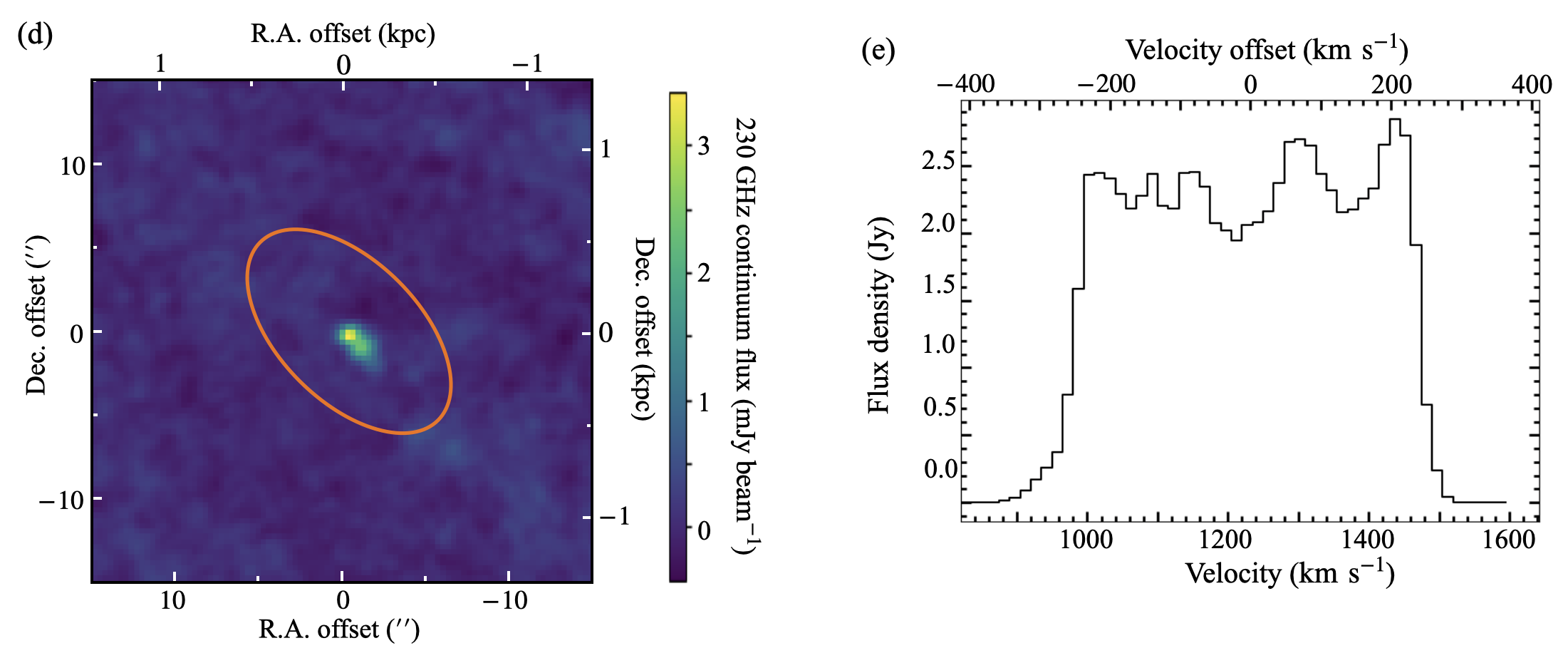}
\caption{\label{fig:CO_maps} Top row: $^{12}$CO(2-1) maps of the central region of NGC~3169 ($\approx3.6\times3.6$~kpc$^2$), encompassing the galaxy bulge, extracted from our ALMA observations as described in Section~\ref{CO}. (a) Moment-0 (integrated intensity) map. (b) Moment-1 (mean velocity) map, with respect to the heliocentric velocity of the galaxy of ($1232$~km~s$^{-1}$; see Section~\ref{target}). (c) Moment-2 (velocity dispersion) map. Bottom row: (d) $230$-GHz continuum emission map. The orange and black ellipses in panels a, b, c and d separate the inner disc and outer star-forming ring discussed in Sections~\ref{analysis}--\ref{discussion}. (e) $^{12}$CO(2-1) spectrum integrated over the whole field of view of the $15$~km~s$^{-1}$ per channel datacube.}
\end{figure*}

The integrated $^{12}$CO(2-1) spectrum and moment-0 (integrated flux), moment-1 (mean velocity) and moment-2 (velocity dispersion) maps are shown in Figure~\ref{fig:CO_maps}. The $230$~GHz continuum map is also shown in Figure~\ref{fig:CO_maps} and reveals a bright region at the galaxy centre, slightly extended along the disc major axis and with an integrated flux of $8.5$~mJy, that could be explained by dust or a jet from the LLAGN. The one-sided structure of this continuum source favors the explanation of a jet, although some VLA radio maps \citep[in e.g.][]{nagar2005radio} show extended structure in the perpendicular direction. Higher resolution observations are needed to resolve the existence and morphology of the jet. Nevertheless, the strong and broadened \ion{N}{ii} emission lines at this location indicate the existence of AGN-driven outflows or a supernova very close to the LLAGN. A recent spectral energy distribution (SED) analysis of the source also reveals a negative spectral index consistent with synchrotron emission (Elford et al. ,in prep). 

The physical scale of the synthesised beam ($99$~pc) allows to estimate the molecular gas mass surface density and thus (combined with H$\alpha$) the SFE at the typical spatial scale of giant molecular clouds (GMCs; $\approx50$~pc; \citeauthor{rosolowsky2003giant} \citeyear{rosolowsky2003giant,rosolowsky2007giant}) and \ion{H}{ii} regions of similar sizes. Stand-alone GMC analyses should however be performed at the highest angular resolution possible for the CO data, here $0\farcs58\times0\farcs58$ ($52$~pc$\times52$~pc), when adopting a more standard Briggs' robust parameter of 0.5 and no taper. 


\section{Surface densities and line ratio}
\label{ini_ana}


\subsection{SFR surface densities}
\label{SFR}

We correct the observed H$\alpha$ fluxes ($F_{\rm H\alpha,obs}$) for extinction using the observed H$\beta$ fluxes ($F_{\rm H\beta,obs}$) and the assumed Balmer decrement described below. Due to differences in the spectral resolutions and observing conditions of the two SITELLE datacubes, the H$\beta$ spaxels that satisfy a $3\sigma$ detection threshold only constitute $52\%$ of the H$\alpha$ spaxels satisfying that same condition. For each spaxel with both the H$\alpha$ and H$\beta$ lines satisfying the $3\sigma$ detection threshold, the "color excess" of H$\alpha$ over H$\beta$ is defined as
\begin{equation}
  E({\rm H}\beta-{\rm H}\alpha)\equiv2.5\log\left(\frac{(F_{\rm H\alpha,obs}/F_{\rm H\beta,obs})}{({\rm H\alpha}/{\rm H\beta})_{\rm init}}\right)~,
\end{equation}
where we assume $({\rm H\alpha}/{\rm H\beta})_{\rm init}=2.86$, as expected for the case B recombination at a temperature of $10^4$~K \citep{osterbrock2006astrophysics}.

The H$\alpha$ extinction is then calculated as
\begin{equation}
  A_{\rm H\alpha}=\left(\frac{E({\rm H\beta}-{\rm H\alpha})}{k(\lambda_{\rm H\beta})-k(\lambda_{\rm H\alpha})}\right)\,k(\lambda_{\rm H\alpha})
\end{equation}
following \citet{nelson2016spatially}, where $k(\lambda)$ is the reddening curve of \citet{fitzpatrick1986average} and $k(\lambda_{\rm H\alpha})$ and $k(\lambda_{\rm H\beta})$ are evaluated at the wavelengths of H$\alpha$ and H$\beta$, respectively. For each spaxel with a $3\sigma$ H$\alpha$ detection but no $3\sigma$ H$\beta$ detection, we adopt the H$\alpha$ extinction of the nearest reliable spaxel. The extinction-corrected H$\alpha$ flux of each spaxel is then obtained as 
\begin{equation}
  F_{\rm H\alpha}=F_{\rm H\alpha,obs}\,e^{A_{\rm H\alpha}/1.086}~.
\end{equation}

This extinction-corrected H$\alpha$ flux is converted to a SFR using the relation of \citet{kennicutt2012star}:
\begin{equation}
  \log({\rm SFR}\,/\,{\rm M}_\odot\,{\rm yr}^{-1})=\log(L_{\rm H\alpha}\,/\,{\rm ergs}\,{\rm s}^{-1})-41.27~,
\end{equation}
where $L_{\rm H\alpha}=F_{\rm H\alpha}(4\pi D^2)$ is the extinction-corrected H$\alpha$ luminosity. We note that when probing spatial scales smaller than $500$~pc, this conversion relation can break down, as seen in examples of \citet{kennicutt2012star}. Local SFR depend on environments and age of the stellar population that we can only estimate here. Nevertheless, this conversion holds true for the radial profile of SFE (see Section~\ref{radialtrend}), which is calculated within apertures large enough for confident H$\alpha-$SFR conversions.

The $\Sigma_{\rm SFR}$ is then calculated as the SFR within each spaxel divided by the spaxel area. The mean $\Sigma_{\rm SFR}$ in the central region of NGC~3169, $\log({\rm \Sigma_{SFR,~2~kpc}~/~M_\odot~yr^{-1}~kpc^{-2}}) = -0.67$, is calculated from the total SFR within a radius of $\approx 1.8$~kpc, as listed in Table~\ref{tab:summary}.

\begin{figure*}
\centering
\includegraphics[width=0.99\textwidth]{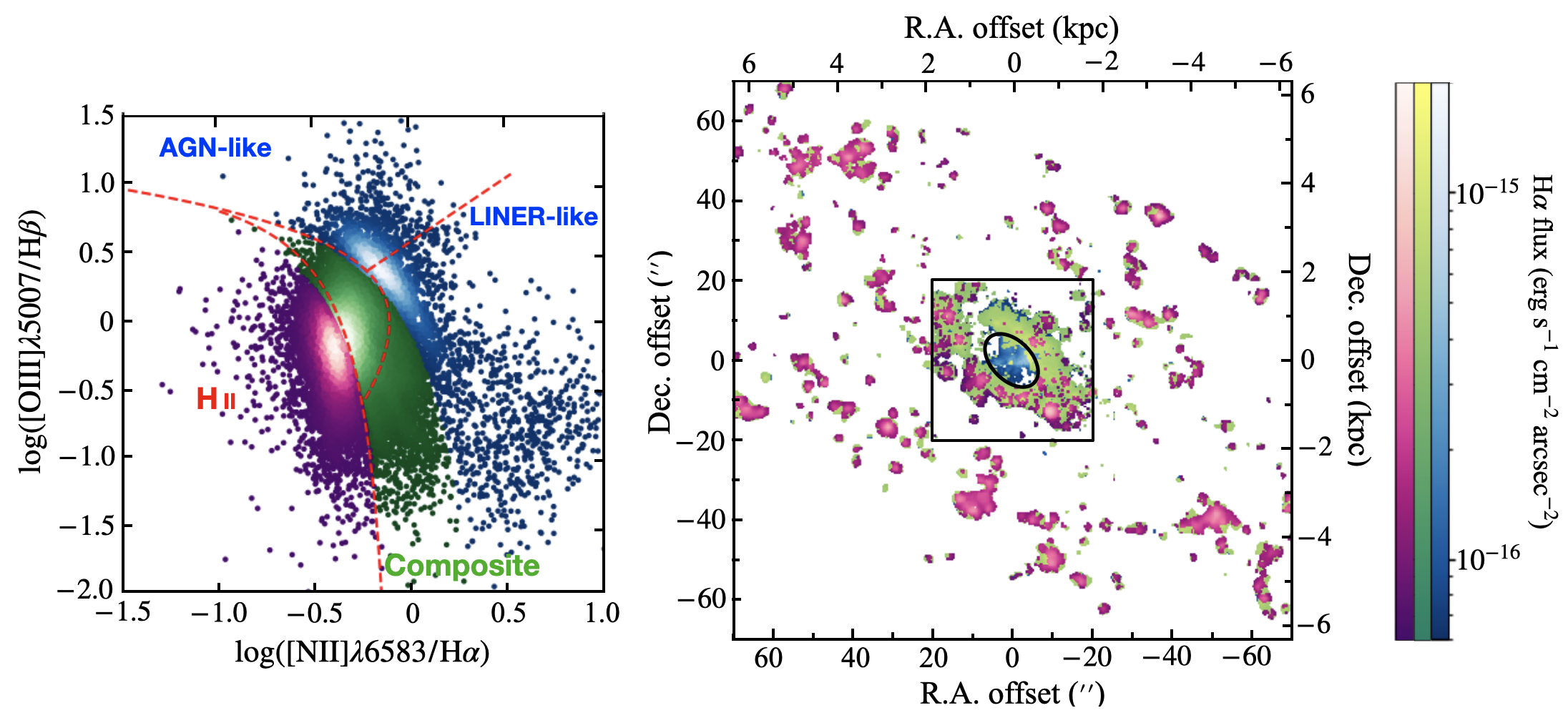}
\includegraphics[width=0.99\textwidth]{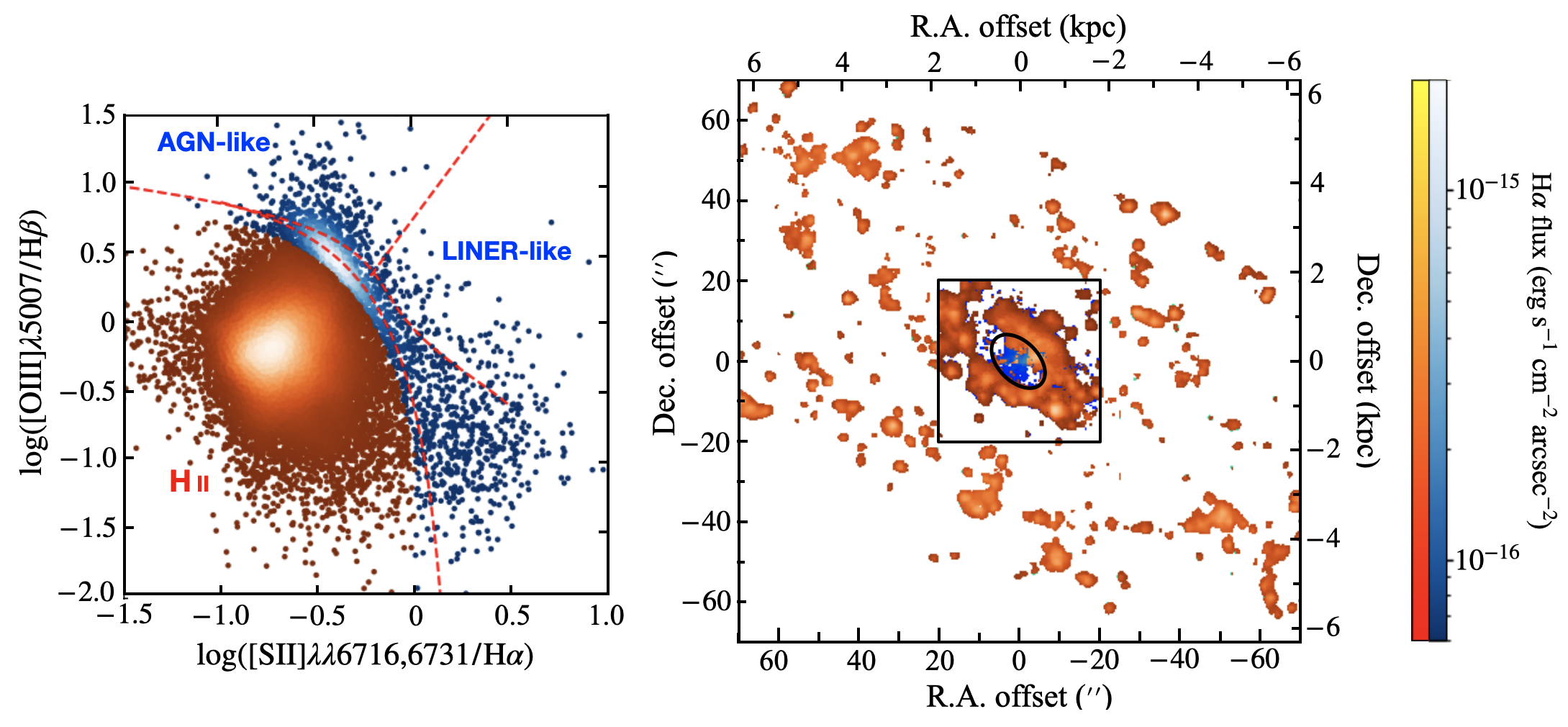}
\caption{\label{fig:BPT_full} Ionisation-mechanism classification over the whole disc of NGC~3169. Top-left: O3N2 diagram, showing the [\ion{O}{iii}]/H$\beta$ versus [\ion{N}{ii}]/H$\alpha$ line ratios. AGN/LINER-ionised (blue), \ion{H}{ii} (purple) and composite (green) regions are separated by the classification criteria of \citet{kewley2006host}. The corresponding regions are colour-coded on the H$\alpha$ flux map in the top-right, indicating the extent of the AGN impact. Bottom-left: O3S2 diagram, showing the [\ion{O}{iii}]/H$\beta$ versus [\ion{S}{ii}]/H$\alpha$ line ratios. AGN/LINER-ionised (blue) and \ion{H}{ii} (orange) regions are again separated by the classification criteria of \citet{kewley2006host}, and the H$\alpha$ flux map in the bottom-right is colour-coded accordingly. The classification lines of \citet{law2021sdss} are overlaid as red lines on the left panels, showing slightly different definitions of the regions and an additional separation of AGN-like and LINER-like ionisation. The black boxes indicates the central $\approx3.6\times3.6$~kpc$^2$ region, encompassing the galaxy bulge. The black ellipses separate the inner disc and outer star-forming ring discussed in Sections~\ref{analysis}--\ref{discussion}.}
\end{figure*}


\subsection{Molecular gas mass surface densities}
\label{gas}

A map of the molecular gas mass surface density ($\Sigma_{\rm H_2}$) can be obtained from the \textsuperscript{12}CO(2-1) observations described in Section~\ref{CO}. First, the CO flux ($F_{\rm CO(2-1)}$) within each spaxel is obtained from the moment-0 map fluxes (in Jy~beam$^{-1}$~km~s$^{-1}$) by dividing by the synthesised beam area in spaxels. $F_{\rm CO(2-1)}$ is then converted to luminosity according to
\begin{equation}
  \frac{L_{\rm CO(2-1)}}{\rm K~km~s^{-1}~pc^2}=\left(\frac{3.25\times10^7}{{(1+z)}^3}\right)\,\left(\frac{F_{\rm CO(2-1)}}{\rm Jy~km~s^{-1}}\right)\,{\left(\frac{\nu_{\rm obs}}{\rm GHz}\right)}^{-2}\,\left(\frac{D}{\rm Mpc}\right)^2
\end{equation}
\citep[e.g.][]{decarli2016alma}, where $z$ is the galaxy redshift and $\nu_{\rm obs}$ is the observed frequency (i.e.\ the redshifted frequency of the CO(2-1) line). 
The luminosity-based molecular gas mass within each spaxel is then calculated using
\begin{equation}
  \frac{M_{\rm H_2}}{{\rm M}_\odot}=4.4\,\left(\frac{L_{\rm CO}}{\rm K~km~s^{-1}~pc^2}\right)\,\left(\frac{X_{\rm CO}}{\rm 2\times10^{20}\,cm^{-2}~(K~km~s^{-1})^{-1}}\right)~,
\end{equation}
where $L_{\rm CO}$ is the CO(1-0) luminosity and $X_{\rm CO}$ is the CO-to-H$_2$ conversion factor. We adopt a CO(2-1)/CO(1-0) ratio of $0.8$ (in brightness temperature units), typical of spiral galaxies \citep[e.g.][]{lamperti2020co}, and $X_{\rm CO}=2.3\times10^{20}$~cm$^{-2}$~(K~km~s$^{-1}$)$^{-1}$. The adopted $X_{\rm CO}$ is commonly used in extragalactic studies \citep[e.g.][]{hughes2013probability,utomo2015giant,sun2018cloud}, although it can depend on the environment of molecular clouds, such as metallicity and radiation field \citep{bolatto2013co}. The mass contribution from helium is considered and included in the coefficient \citep{strong1988radial,bolatto2013co}. The complete conversion thus becomes
\begin{equation}
  \frac{M_{\rm H_2}}{{\rm M}_\odot}=6.325\,\left(\frac{L_{\rm CO(2-1)}}{\rm K~km~s^{-1}~pc^2}\right).
\end{equation}
The molecular gas mass surface density is then calculated as $M_{\rm H_2}$ divided by the spaxel area. The mean $\Sigma_{\rm H_2}$ in the central region of NGC~3169, $\log({\rm \Sigma_{\rm H_2,~2~kpc}~/~M_\odot~pc^{-2}}) = 2.31$, is calculated from the total molecular gas mass within a radius of $\approx 1.8$~kpc, as listed in Table~\ref{tab:summary}.


\subsection{Emission-line ratios}
\label{line_ratio}

Considering the existence of the LLAGN at the centre of NGC~3169, it is unclear what fraction of the H$\alpha$ emission originates from star formation, especially within a $1$~kpc radius of the galaxy centre. To accurately determine the SFEs in this region, we must distinguish the different ionisation mechanisms and confidently identify the \ion{H}{ii} regions. Here, we utilise Baldwin, Phillips $\&$ Terlevich (BPT) diagrams \citep{baldwin1981classification} to assess the extent of the AGN impact. 
We consider the ionised-gas emission-line ratios of [\ion{O}{iii}]/H$\beta$, [\ion{N}{ii}]/H$\alpha$ and [\ion{S}{ii}]/H$\alpha$. These ratios are most relevant when distinguishing \ion{H}{ii} regions from AGN-ionised regions \citep[as described in][]{baldwin1981classification} and they are readily available from our SITELLE datacubes. Figure~\ref{fig:BPT_full} shows two different BPT diagrams: the "O3N2" diagram, which compares the [\ion{N}{ii}]/H$\alpha$ ratios with [\ion{O}{iii}]/H$\beta$, and the "O3S2" diagram, which compares the [\ion{S}{ii}]/H$\alpha$ ratios with [\ion{O}{iii}]/H$\beta$. We applied the classification lines of \citet{kewley2006host} to separate and thus uniquely identify \ion{H}{ii}, AGN-ionised and composite regions using the O3N2 diagram, and to separate \ion{H}{ii} and Seyfert/LINER-ionised regions using the O3S2 diagram. Spaxels are colour-coded accordingly to these ionisation mechanisms in the flux maps shown in Figure~\ref{fig:BPT_full}. The classification lines of \citet{law2021sdss} are overlaid on the left panels of Figure~\ref{fig:BPT_full} for reference, as they additionally show the separation of AGN-like and Seyfert-like regions.

From the O3S2 diagram, the majority of spaxels can be identified as belonging to \ion{H}{ii} regions, except within the central $650$~pc radius where only a small fraction of the spaxels contain gas definitely ionised by star formation. On the other hand, the O3N2 diagram shows more ambiguous results. While the spaxels within the central $650$~pc radius also appear to be affected by AGN ionisation, the composite regions are much more extended, encompassing $\approx50\%$ of the spaxels in the central region and reaching beyond a radius of $2$~kpc from the galaxy centre. The exact fraction of star-formation ionisation in these composite and AGN-ionised regions is thus unclear. Nevertheless, it is suggested from the maps that AGN-ionisation occupies the diffuse emissions surrounding H$\alpha$ emission peaks. 

The SFE analyses in Sections~\ref{analysis} and \ref{discussion} thus systematically consider the three cases based on the O3N2 diagram separately. Spaxels belonging to \ion{H}{ii} regions (colour-coded purple) are referred to as star-forming regions and their SFRs can be confidently derived from their H$\alpha$ emissions. For spaxels that occupy composite regions (colour-coded green) and AGN-ionised regions (colour-coded blue), upper limits to the SFRs are estimated. The distinction between AGN-like and LINER-like ionisation mechanisms is unclear, due to insufficient $S/N$ of the H$\beta$ and \ion{O}{iii} lines. Nevertheless, the majority of these regions can be classified as AGN-like based on the O3S2 diagram, supporting the hypothesis that AGN photoionisation is the dominant ionisation mechanism within the inner $1$~kpc radius of NGC~3169. However, there are also a few LINER-like regions presumably due to old stellar populations. 

We also note that recent studies \citep[e.g.][]{law2021sdss} proposed using 3D diagnostic diagrams in the parameter space of [\ion{O}{iii}]/H$\beta$, [\ion{N}{ii}]/H$\alpha$ and [\ion{S}{ii}]/H$\alpha$, to address the difference between the O3S2 and O3N2 diagrams. We include this alternative analysis in Appendix \ref{app_3DBPT}, which confirms the nature of \ion{H}{ii} regions while showing that the composite regions are likely AGN-ionised.


\section{Analysis: depletion time and gas dynamics}
\label{analysis}


\subsection{Spatially-resolved depletion times}
\label{depttime_map}

\begin{figure}
\centering
\includegraphics[width=0.49\textwidth]{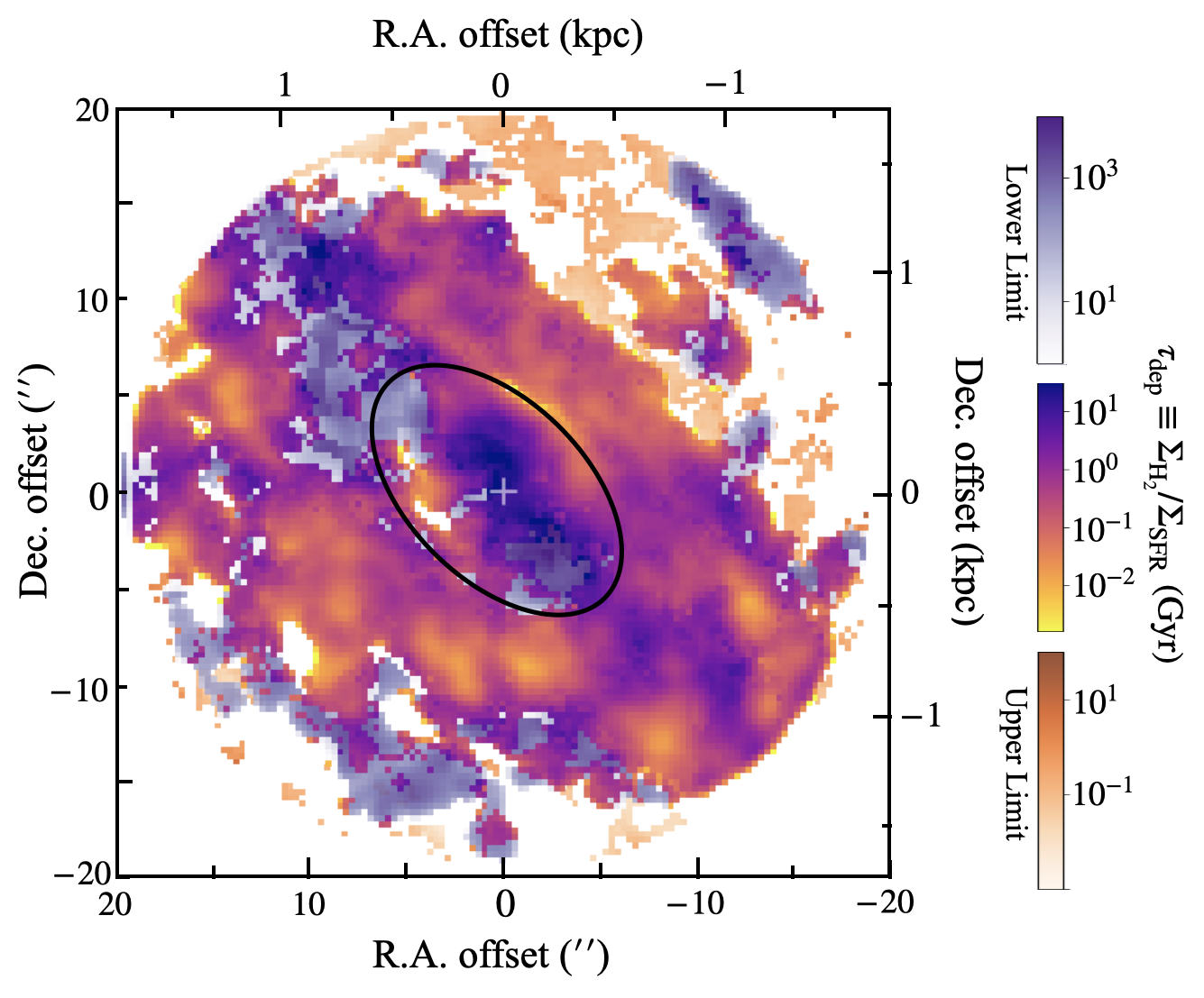}
\caption{\label{fig:dept_map}Depletion time map of the central region of NGC~3169 ($\approx3.6\times3.6$~kpc$^2$), encompassing the galaxy bulge, calculated from the $\Sigma_{\rm H_2}$ and $\Sigma_{\rm SFR}$ maps (see Section~\ref{depttime_map}). The orange-purple colour scale shows actual measurements, for spaxels where both CO and H$\alpha$ are above our detection threshold. The light-orange color scale shows upper limits, for spaxels where only H$\alpha$ is above our detection threshold. The grey-purple color scale shows lower limits, for spaxels where only CO is above our detection threshold. Blank spaxels are either outside the ALMA field of view or have both H$\alpha$ and CO below our detection threshold. The black ellipse separates the inner disc and outer star-forming ring discussed in Sections~\ref{analysis}--\ref{discussion}.}
\end{figure}

Figure~\ref{fig:dept_map} shows the spatially-resolved depletion time map, constructed from the ratio of the SFR and molecular gas mass surface densities ($\tau_{\rm dep}\equiv\Sigma_{\rm H_2}/\Sigma_{\rm SFR}$). For spaxels where both $\Sigma_{\rm SFR}$ and $\Sigma_{\rm H_2}$ are reliably measured (i.e.\ where both H$\alpha$ and CO are detected with $S/N>3$), $\tau_{\rm dep}$ is calculated by simply dividing the two surface densities. This ratio is shown in Figure~\ref{fig:dept_map} with an orange-purple colour scale. However, many spaxels have only one tracer brighter than our adopted detection threshold. When only $\Sigma_{\rm SFR}$ is reliably measured, we estimate a $\tau_{\rm dep}$ upper limit by dividing the $3\sigma$ upper limit on the molecular gas mass surface density by the reliably measured SFR surface density, shown in Figure~\ref{fig:dept_map} with a light-orange corlour scale. An analogous approach is applied to spaxels where only $\Sigma_{\rm H_2}$ is reliably measured, to estimate a $\tau_{\rm dep}$ lower limit, shown in Figure~\ref{fig:dept_map} with a grey-purple colour scale.

We note that Figure~\ref{fig:dept_map} purposefully shows all spaxels where a tracer is detected, without distinguishing them based on the dominant ionisation mechanism. For spaxels identified as composite or AGN-ionised based on their emission-line ratios (see Section~\ref{line_ratio}), the SFR is therefore likely an upper limit. In turn, the derived depletion times measurements are likely lower limits. We return to this issue in Sections~\ref{KS} and \ref{radialtrend}.

The region of NGC~3169 within the ALMA FOV has a radius of $\approx1.8$~kpc, and will be referred to here as the "central region". It is nearly completely filled with molecular gas, while the ionised-gas emission is more patchy and concentrated within a ring-like structure encompassing radii of $\approx0.65$~kpc to $\approx1.8$~kpc. The average depletion time within the ring is $\approx0.3$~Gyr, calculated from the ratio of total molecular gas mass to total SFR. It is consistent with typical extragalactic starburst regions \citep[e.g.][]{barrera2021self}. The ring may constitute a resonance ring \citep[e.g.][]{1996FCPh...17...95B,buta1998ngc,comeron2014arrakis} or may simply be a pseudo-ring formed by winding spiral arms. Inside the ring (radii $\lesssim650$~pc), the molecular gas is more concentrated and star formation appears to be suppressed. This inner disk also has stronger impact from the bulge, with B/D$>1$. The radial variation of the depletion time is discussed in Section~\ref{radialtrend}.


\begin{figure*}
\centering
\includegraphics[width=0.99\textwidth]{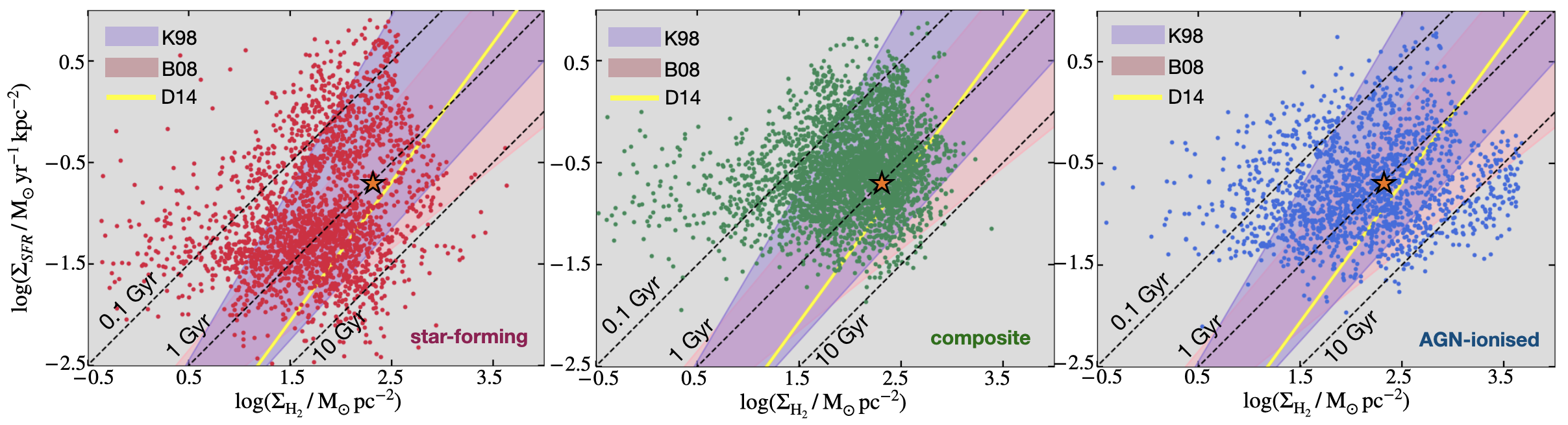}
\caption{\label{fig:dept_KS} Plots of $\Sigma_{\rm H_2}$-$\Sigma_{\rm SFR}$, overlaid with the power-law relations of \citet{kennicutt1998global} and \citet{bigiel2008star}, shown as light-purple and light-pink shaded regions, respectively. The power-law relation of \citet{davis2014atlas3d} for early-type galaxies is shown as a yellow line. Using the BPT analysis of Section~\ref{line_ratio}, from left to right the panels include only spaxels classified as star-forming, composite and AGN-ionised, respectively. All data points are from spaxels with both H$\alpha$ and CO above our detection threshold. The mean $\Sigma_{\rm H_2}$-$\Sigma_{\rm SFR}$ of within the ALMA FOV is shown as an orange star in each panel. Representative depletion timescales of $0.1$, $1$ and $10$~Gyr are shown as black dashed lines.}
\end{figure*}

\subsection{SFR to molecular gas mass relation}
\label{KS}

Figure~\ref{fig:dept_KS} shows the $\Sigma_{\rm SFR}$-$\Sigma_{\rm H_2}$ relation of all the spaxels with reliably-detected H$\alpha$ and CO (orange-purple colour scale in Figure~\ref{fig:dept_map}). However, contrary to Figure~\ref{fig:dept_map}, we separate and compare the three cases discussed in Section~\ref{line_ratio}: star-forming (orange scale), composite (green scale) and AGN-ionised (blue scale) spaxels, based on the O3N2 type BPT diagnostic described in Section~\ref{line_ratio}. It is important to recall that for the composite and AGN-ionised cases, $\Sigma_{\rm SFR}$ has large uncertainties and should really be treated as an upper limit. The mean $\Sigma_{\rm SFR}$ and $\Sigma_{\rm H_2}$ within the ALMA field of view is shown with an orange star in Figure~\ref{fig:dept_map}. The distributions are compared to characteristic timescales ($0.1$, $1$ and $10$~Gyr) and the power-law relations of \citet{kennicutt1998global} and \citet{bigiel2008star}, shown as light-purple and light-pink shaded areas, respectively. The power-law relation of \citet{davis2014atlas3d} for early-type galaxies is also shown as yellow lines for reference.

Because ionised-gas and molecular gas peaks are generally not exactly co-spatial \citep[e.g.][]{schruba2010scale,schinnerer2019physics,2019Natur.569..519K,pan2022gas}, it is common to have broad $\Sigma_{\rm SFR}$-$\Sigma_{\rm H_2}$ distributions at sub-kpc scales. In fact, this scatter can be interpreted as a sign of a rapid cycling between gas and stars on the sub-kpc scales, which allows us to constrain molecular gas lifetimes, as shown in Section~\ref{dis:tuning}.  Figure~\ref{fig:dept_KS} reveals that this is indeed the case in NGC~3169, but the $\Sigma_{\rm SFR}$-$\Sigma_{\rm H_2}$ measurements roughly follow the \citet{kennicutt1998global} and \citet{bigiel2008star} relations in each case, encompassing depletion timescales of $0.1$ to $10$~Gyr. Nevertheless, the three distributions peak at different $\Sigma_{\rm SFR}$ and $\Sigma_{\rm H_2}$. Although some star-forming spaxels have short depletion timescales (see left panel of Fig.~\ref{fig:dept_KS}), many star-forming spaxels occupy a part of the $\Sigma_{\rm SFR}$-$\Sigma_{\rm H_2}$ space centred at $\Sigma_{\rm SFR}\approx0.04$~M$_\odot$~yr$^{-1}$~kpc$^{-2}$ and $\Sigma_{\rm H_2}\approx100$~M$_\odot$~pc$^{-2}$ ($\tau_{\rm dep}\approx2.5$~Gyr), slightly larger than the average $\tau_{\rm dep}\approx2$~Gyr reported by \citet{bigiel2008star}. 
In the composite regions, shown in the middle panel of Figure~\ref{fig:dept_KS}, the distribution is centred at $\Sigma_{\rm SFR}\approx0.3$~M$_\odot$~yr$^{-1}$~kpc$^{-2}$ and $\Sigma_{\rm H_2}\approx160$~M$_\odot$~pc$^{-2}$ ($\tau_{\rm dep}\approx0.5$~Gyr). However, this results from ionised-gas emission that is only partly due to star formation (i.e.\ the derived $\Sigma_{\rm SFR}$ of composite regions are really upper limits), so this peak would be lowered in the figure (higher $\tau_{\rm dep}$) if we knew the exact fraction of the star-forming radiation. As shown in the right panel of Figure~\ref{fig:dept_KS}, the $\Sigma_{\rm SFR}$-$\Sigma_{\rm H_2}$ distribution of AGN-ionised regions is similar to that of composite regions, but more scattered and extended.


\subsection{Radial trend of depletion time}
\label{radialtrend}

\begin{figure}
\centering
\includegraphics[width=0.42\textwidth]{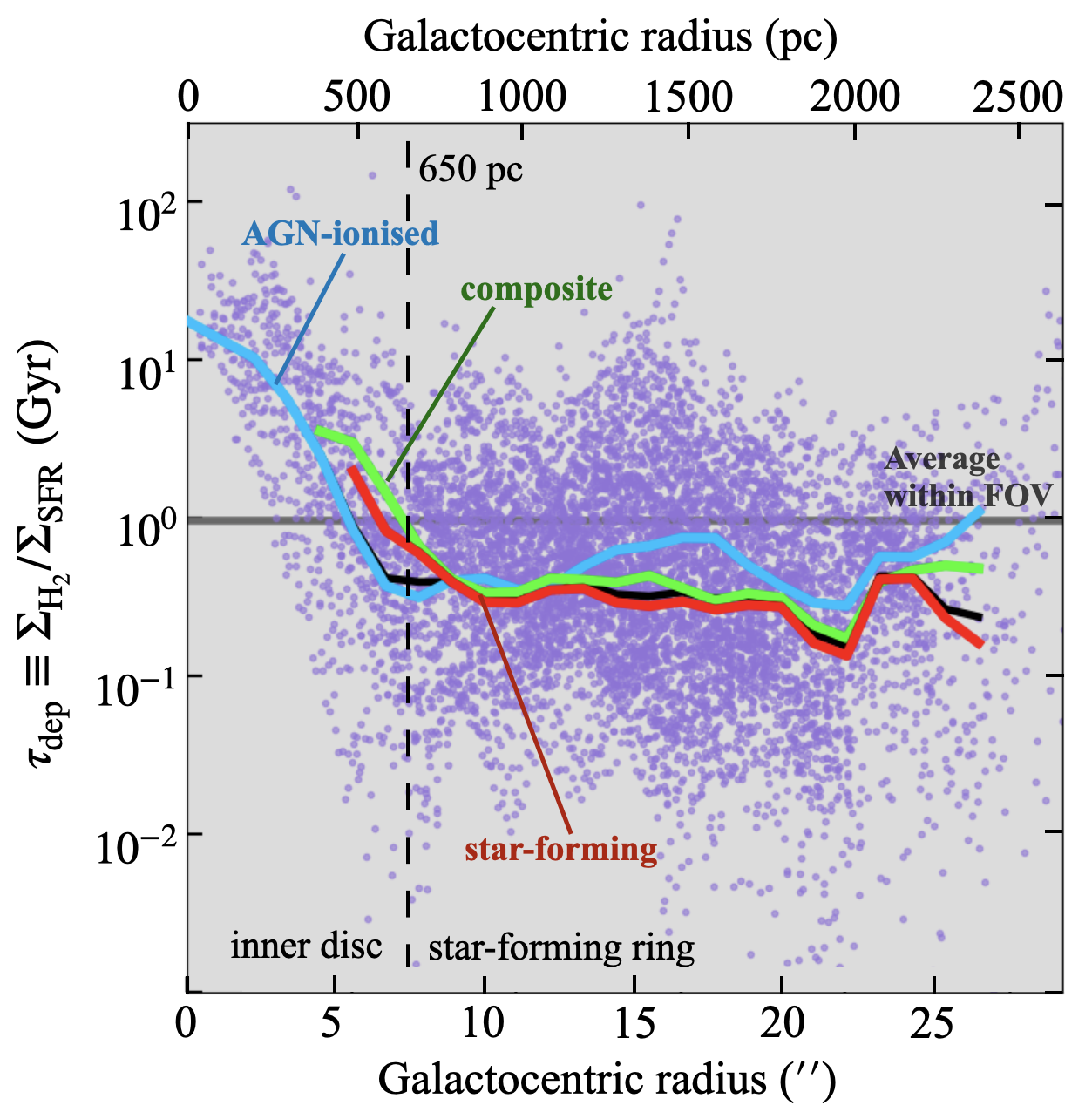}
\caption{\label{fig:dept_rad} Depletion time radial profile in the central region of NGC~3169, encompassing the galaxy bulge. Purple datapoints are calculated from all the spaxels with both H$\alpha$ and CO above our detection threshold in Figure~\ref{fig:dept_map}. The black line is the running mean of the depletion times of all datapoints as a function of galatocentric radius. The red, green and blue lines are the running means of the star-forming, composite and AGN-ionised regions only, respectively. The vertical black dashed line separates the inner disc and outer star-forming ring at a radius of $650$~pc discussed in Sections~\ref{analysis}--\ref{discussion}. The grey horizontal line shows the mean depletion time within the ALMA FOV, corresponding to the orange star in Figure~\ref{fig:dept_KS}.}
\end{figure}

Figure~\ref{fig:dept_rad} shows $\tau_{\rm dep}$ as a function of deprojected galactocentric radius, where individual spaxels are shown as purple datapoints, and the running mean for each case discussed in Section~\ref{line_ratio} is shown in the associated colour, using annuli of width $200$~pc and a step size of $100$~pc. 
Only spaxels reliably-detected in both H$\alpha$ and CO are included in this figure. For reference, the global depletion time ($0.9$ Gyr) is indicated by a grey horizontal line, calculated from the ratio of the total SFR to the total molecular gas mass within a radius of $\approx1.8$~kpc that is the ALMA FOV.

Between galactocentric radii of $0.65$ and $1.8$~kpc, i.e.\ within the previously-discussed star-forming ring, the depletion times of individual spaxels cover a wide temporal range ($0.03$ -- $10$~Gyr). The mean $\tau_{\rm dep}$ within this region is $\approx0.3$~Gyr, lower than the global $\tau_{\rm dep}$ of $0.9$~Gyr. Larger variations of the average depletion times occur beyond a radius of $1.8$~kpc, marking the outer edge of the star-formation ring. As shown in Figure~\ref{fig:dept_map}, GMCs and ionised-gas clouds are particularly not co-spatial beyond this radius.

Within a galactocentric radius of $\approx0.9$~kpc, the average $\tau_{\rm dep}$ rapidly increases with decreasing radius, with or without the inclusion of composite and AGN-ionised regions. For exclusively star-forming regions (orange curve in Figure~\ref{fig:dept_rad}), the mean $\tau_{\rm dep}$ reaches $\approx2.3$~Gyr at a radius of $\approx500$~pc. The further inclusion of composite and AGN-ionised regions (blue curve) is considered merely for illustration purposes, but clearly shows that there is H$\alpha$ emission down to a radius of $\approx100$~pc, partly as the result of AGN ionisation. The measurements nevertheless clearly indicate depletion times of at least the age of the Universe (most likely more) in the central few hundred parsecs. Therefore, although the average $\Sigma_{\rm SFR}$ and $\Sigma_{\rm H_2}$ of this central region follows the SF relations of star-forming regions, we observe a clear transition from efficient SF in the star-forming ring to quenched SF within the central $\approx500$~pc in radius. These low SFEs result from feeble star formation in that region, despite plentiful molecular gas, potentially due to accretion onto the central regions.

The radial extent of AGN ionisation is also illustrated in Figure~\ref{fig:dept_rad}. While H$\alpha$ emissions reaches the centre of the galaxy, the inner $400$~pc in radius are dominated by AGN ionisation. Individual spaxels that are confidently classified as exclusively star-forming are located at a galactocentric radius of at least $650$~pc. The size of the AGN-ionised region is consistent with that of other nearby galaxies. For example, it is within the broad range of $0.3$--$10$~kpc reported by \citet{chen2019spatial} based on 152 AGN-host galaxies using the Mapping Nearby Galaxies at Apache Point Observatory (MaNGA) survey. 


\subsection{Virial parameters}
\label{virial}

To take a more in-depth look at the star formation on cloud scales, we identified \ion{H}{ii} regions using the SITELLE H$\alpha$ data and algorithms developed by \citet{rousseau2018ngc628}, summarised below. The H$\alpha$ flux peaks are first identified, based on the following criteria: (1) the flux of the peak spaxel is greater than the flux of at least five immediately surrounding spaxels, (2) the total flux in a $3\times3$~spaxels box centred on the emission peak is above the $3\sigma$ detection threshold and (3) if two emission peaks are separated by a distance smaller than the seeing ($1\farcs1$ or $99$~pc in this case), only the brightest peak is preserved. Given an estimate of the characteristic radius of \ion{H}{ii} regions, spaxels are then classified into zones surrounding the peaks, and the boundaries of the spatial zones are defined. Figure~\ref{fig:HII_app} in Appendix~\ref{app_HIIregion} shows these peaks and region boundaries overlaid on the H$\alpha$ flux map of NGC~3169. For each \ion{H}{ii} region, a two-dimensional (2D) Gaussian fit is then carried out and the luminosity ($L_{\rm \ion{H}{ii}}$ based on H$\alpha$ flux) and size (half-width at half-maximum $R_{\rm \ion{H}{ii}}$) of the best-fitting Gaussian are calculated (and taken as those of the \ion{H}{ii} region). Furthermore, the spectra from the SITELLE H$\alpha$ datacube are binned around the best-fitting peak of each \ion{H}{ii} region, according to the size of the region, and the resulting spectrum is fit with {\tt ORCS} to derive the velocity dispersion ($\sigma_{\rm \ion{H}{ii}}$) of the region. We note that not all \ion{H}{ii} regions identified are necessarily also star-forming regions, as some may be partially or primarily ionised by the AGN. The full \ion{H}{ii} region catalogue is presented in Table~\ref{tab:HII} in Appendix~\ref{app_HIIregion}. 

Based on these measurements, the virial parameter of each region is calculated as
\begin{equation}
  \alpha_{\rm vir,\ion{H}{ii}}\equiv5\,\frac{\sigma_{\rm \ion{H}{ii}}^2R_{\rm \ion{H}{ii}}}{GM_{\rm \ion{H}{ii}}}
\end{equation}
\citep{bertoldi1992pressure}, where $M_{\rm \ion{H}{ii}}$ is the ionised-gas mass of the region, estimated from
\begin{equation}
  \frac{M_{\rm \ion{H}{ii}}}{\rm M_\odot}=1.57\times10^{-17}\,\left(\frac{L_{\rm \ion{H}{ii}}}{\rm erg~s^{-1}}\right)^\frac{1}{2}\,\left(\frac{R_{\rm \ion{H}{ii}}}{\rm pc}\right)^\frac{3}{2}  
\end{equation}
\citep{zaragoza2015comparative}. This ionised-gas mass assumes that the mean electron number volume density can be derived from $L_{\rm \ion{H}{ii}}$ assuming a spherical geometry, and that all ionising photons are from hot/massive stars.

\begin{figure*}
\centering
\includegraphics[width=0.9\textwidth]{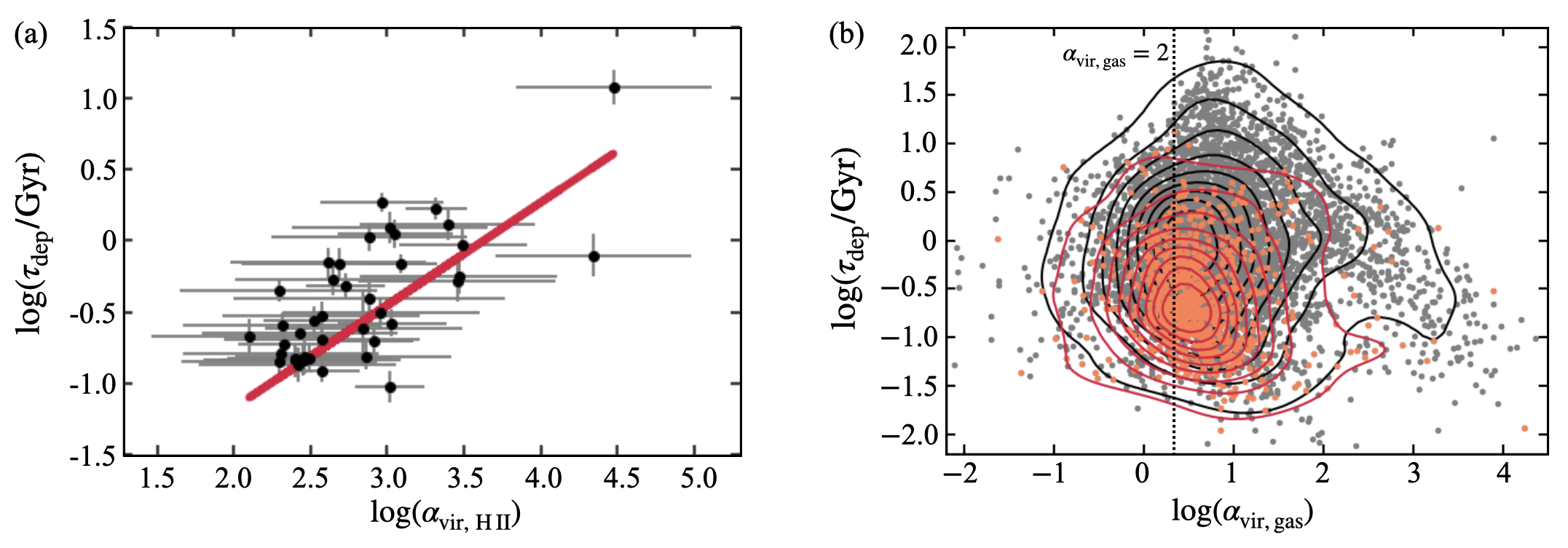}
\caption{\label{fig:virial} 
(a) \ion{H}{ii} region depletion time versus virial parameter, with the best-fitting power-law relation overlaid as a red line. (b) Depletion time versus virial parameter calculated from the molecular gas at each spaxel. The orange datapoints and associated red density contours are based on spaxels classified as star-forming only, while the grey datapoints and associated black density contours are based on all the spaxels with reliable measurements of $\tau_{\rm dep}$ and $\alpha_{\rm vir,\,gas}$.}
\end{figure*}

As the ionised-gas mass is usually only a fraction of the total gas mass \citep{zaragoza2015comparative}, the virial parameter calculated above is typically larger than that based on molecular gas. Therefore, it is not fair to use a criterion of about unity to diagnose whether a cloud is gravitationally bound (often stated as $\alpha_{\rm vir}\lesssim2$). However, it is meaningful to compare virial parameters measured for different \ion{H}{ii} regions. A larger virial parameter implies a higher kinetic energy, that is harder to counter-balance with gravity and is more likely to require external pressure to maintain equilibrium. As should then be expected, there is a weak correlation between $\alpha_{\rm vir,\ion{H}{ii}}$ and $\tau_{\rm dep}$, shown in the left panel of Figure~\ref{fig:virial}. The $\tau_{\rm dep}$ in each \ion{H}{ii} region is calculated based on the ratio of molecular gas mass and SFR within the region. According to equations (8), variations of $\alpha_{\rm vir,\ion{H}{ii}}$ are in fact primarily due to the velocity dispersions. This suggests that large $\tau_{\rm dep}$ in the inner $1$~kpc radius can be associated with high turbulence.

As a comparison, we also estimate the molecular gas virial parameter on the cloud scale, which is equivalent to the CO spaxel size of $\approx35$~pc. The molecular gas virial parameter ($\alpha_{\rm vir,\,H_2}$) can be calculated following the method from \citet{rosolowsky2006bias} that is recently adopted by \citet{sun2020molecular}:
\begin{equation}
    \alpha_{\rm vir,\,H_2}=3.1\times\,\left(\frac{\Sigma_{\rm H_2}}{\rm 10^2\,M_\odot\,{\rm pc}^2}\right)^{-1}\left(\frac{\sigma_{\rm H_2}}{\rm 10\,km\,s^{-1}}\right)^2\left(\frac{D_{\rm spaxel}}{\rm 150\,pc}\right)~,
\end{equation}
where $\Sigma_{\rm H_2}$ is calculated in Section~\ref{gas}, $\sigma_{\rm H_2}$ is taken from the moment-2 map shown in Figure~\ref{fig:CO_maps} and $D_{\rm spaxel}$ is the CO spaxel size of $\approx35$~pc, comparable to typical GMC sizes. As shown in the right panel of Figure~\ref{fig:virial} with orange datapoints and associated red distribution density contours, the majority of spaxels belonging to star-forming regions appear to be roughly virialised, with virial parameters varying between $0$ and $10$ and depletion times less than $1$~Gyr. A virial parameter equals 2 is shown with a vertical dotted line on the right panel of Figure~\ref{fig:virial}, for illustration purposes. On the other hand, a wider range of $\alpha_{\rm vir,\,gas}-\tau_{\rm dep}$ parameter space is covered by all reliably measured spaxels (i.e. including those with composite and AGN ionisation), shown with grey datapoints and associated black distribution density contours in the right panel of Figure~\ref{fig:virial}. This suggests that AGN-ionisation increases both $\alpha_{\rm vir,\,gas}$ and $\tau_{\rm dep}$. However, none of the mean molecular mass surface density, velocity dispersion and/or size is accurately estimated from that within a single spaxel, so the virial parameters evaluated using equation (10) should not be taken nor interpreted literally. Instead, they are rough indicators of the significance of the kinetic energy (turbulence) relative to the gravitational potential (gravity), which varies among spaxels.


\subsection{Non-circular motions}
\label{vfield}

We now probe both molecular gas and ionised gas kinematics in the central region of NGC~3169, to understand more about the SFEs in the inner disc and the star-forming ring. 

\begin{figure*}
  \centering
  \includegraphics[width=0.98\textwidth]{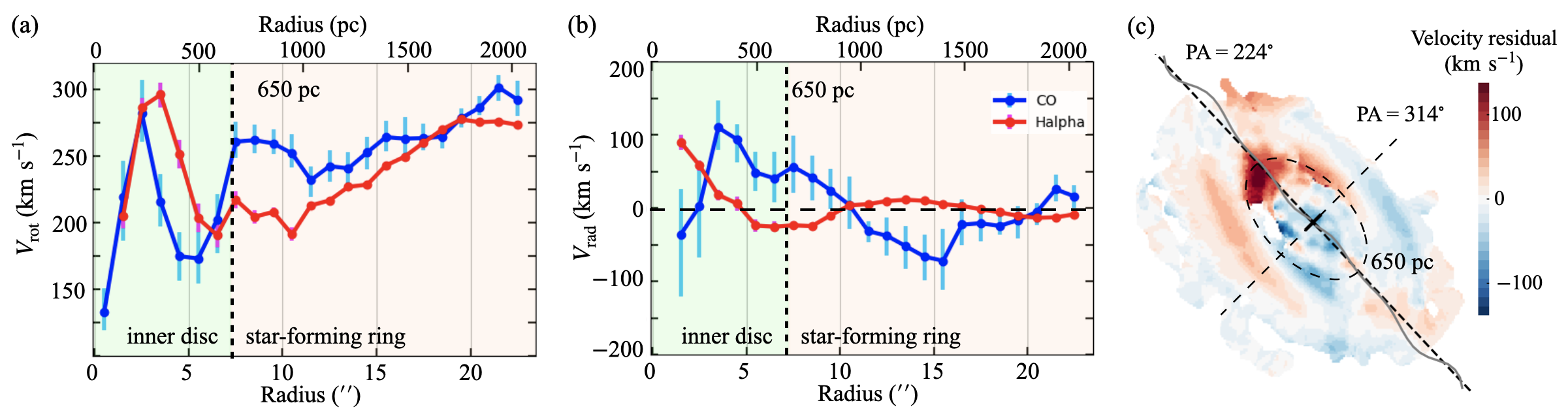}
  \caption{\label{fig:vrot_rad}(a) Rotation curve (left) and radial velocity profile (right) for both CO (blue line) and H$\alpha$ (red line) gas. The green and orange backgrounds are separated at the dotted black line, indicating a radius of $650$~pc that is the intersection of the inner disc and the star-forming ring. (b) CO velocity residual map, defined as the difference between the observed velocity, shown in Figure~\ref{fig:CO_maps}b, and the best-fitting circular velocity model from {\tt 3DBarolo}.}
\end{figure*}

For the molecular gas, we use the {\tt 3DBarolo} software\footnote{https://editeodoro.github.io/Bbarolo/} of \citet{teodoro20153d}, that fits tilted-ring models to three-dimensional (3D) emission-line datacubes, taking advantage of the high spectral and spatial resolutions of the ALMA CO observations. Following a first fit with all parameters free, we fix the dynamical centre (spatially and spectrally)and inclination, while letting the position angle (PA) vary to account for warping of the disc. The rotation and radial velocities from the resulting best-fitting model are shown in Figure~\ref{fig:vrot_rad} with blue lines. As a reference, the corresponding position-velocity (PV) profile and best-fitting reults of all the relevant parameters are in Appendix~\ref{app_CO_Vrad}. 

We use another method to determine the rotation and radial velocity profiles of the ionised gas, by performing Markov Chain Monte Carlo (MCMC) fits of the tangential and radial velocity components. A nuisance parameter is also applied, which corresponds to the line-of-sight velocity scatter in the axi-symmetric modelling of the observed asymmetric kinematics. The Bayesian inferences are done using the Python library \textsc{emcee} \citep{foreman2013emcee}, fixing the dynamical centre, inclination and PA of the H$\alpha$ velocity field as the best-fitting values obtained from the 3DBarolo for the CO gas. This way, we use the same set of kinematic parameters for both gaseous components to assist the comparisons, while taking into account the individual errors on the line-of-sight H$\alpha$ velocities, which are unique to the SITELLE datacubes. The best-fitting rotation and radial velocities are shown in Figure~\ref{fig:vrot_rad} with red lines.

The rotation and radial velocity profiles of CO and H$\alpha$ gas show different behaviors within the inner disc and in the star-forming ring, marked in Figure~\ref{fig:vrot_rad} with green and orange shaded areas, respectively. The inner edge of the star-forming ring at $650$~pc radius is marked with a black dotted line. Beyond the radius of $\approx650$~pc, the derived rotation curve of CO gas suggests roughly constant rotation velocities of $240$--$260$~km~s$^{-1}$ in the star-forming ring. The rotation curve of H$\alpha$ gas, however, increases from $200$~km~s$^{-1}$ at $650$~pc to $280$~km~s$^{-1}$ at $1700$~pc. 

Higher CO rotation velocities are found in many galaxies \citep{davis2013atlas3d,levy2018edge}. If H$\alpha$ is tracing recent star formation, it would typically co-rotate with CO gas in the mid-plane of the galaxy disc. The discrepancy in NGC~3169 can be explained by a different scale height of molecular and ionised gas, due to the fact that the H$\alpha$ gas is associated with a mixture of ionisation mechanisms, including old stellar population and shock-ionisation from the AGN. Based on our ionised gas emission line ratio analysis in Section~\ref{line_ratio}, this is likely the case. The positive slope of H$\alpha$ rotation velocities from inner parts of the star-forming ring to outer parts, where they merge with CO rotation velocities, also indicate that the fraction of recent star-forming ionisation is dominant as we move away from the centre of the galaxy.

The kinematics in the inner disc appear to be much more complex. Both CO and H$\alpha$ rotation velocities rise to $\approx300$~km~s$^{-1}$ and then rapidly drop to below $200$~km~s$^{-1}$. The resulting rotation curve resembles those of many barred galaxies, with a characteristic ``double-hump'' (i.e., a steep inner rise of the rotation velocities, followed by a decrease and local minimum before a second increase to the flat part of the rotation curve; e.g.\ \citealt{bureau1999nature, athanassoula1999bar}). Although NGC~3169 does not harbour a large-scale bar, Figs.~\ref{fig:HST} and \ref{fig:CO_maps} suggest a small-scale bar within the star-forming ring (half-length $\lesssim650$~pc). This structure may be responsible for transporting material to the inner regions, causing high turbulence. 

As shown in the middle panel of Figure~\ref{fig:vrot_rad}, the radial velocities of the CO and H$\alpha$ gas oscillate from positive to negative, tracing expansion and contraction at different locations. In particular, the oscillations of CO and H$\alpha$ are out of phase, indicating complex non-circular motions. The right panel of Figure~\ref{fig:vrot_rad} shows the residual velocity of CO gas, which is the difference between observed velocity (as shown in Figure~\ref{fig:CO_maps}~) and the best-fit model from {\tt 3DBarolo}. It shows the presence of significant spiral arm-like residuals. Given the presence of a small-scale bar, such residuals may trace inflow/outflow streaming motions induced by the inner bar \citep[see e.g.][]{dominguez2020searching,ruffa2022agn}. Another source of non-circular motion is outflows. In Figure~\ref{fig:vrot_rad}~(c), there appear to be particularly large CO residual velocities near the intersection of the major axis and the $650$~pc line. At this location, we also find secondary emission peaks on the CO spectra, as well as asymmetric broadening of ionized gas emission lines. As these features are close to the galaxy centre, they suggest the presence of gas blobs ejected from the galaxy plane, as a result of AGN-driven outflows. However, the signal-to-noise at this location is not robust enough to draw concrete conclusions.

Regardless of the sources of non-circular motions, high velocity dispersions are observed within the inner disc as a result. As shown in Figure~\ref{fig:vel_field_app}, the molecular gas velocity dispersions are high throughout ($\sigma_{\rm gas}\gtrsim20$~km~s$^{-1}$) and rise rapidly with decreasing radius within the inner disc, reaching $\approx70$~km~s$^{-1}$ in the very centre. Since {\tt 3DBarolo} takes into consideration beam smearing effects, the best-fitting velocity dispersion result can assist the moment-2 maps in Figure~\ref{fig:CO_maps} to illustrate the turbulent gas motions within the inner region of NGC~3169.


\subsection{Molecular cloud timescales}
\label{dis:tuning}

A spatial de-correlation between CO and H$\alpha$ emission regions is commonly observed in nearby galaxies \citep[e.g.][]{2019Natur.569..519K,2020MNRAS.493.2872C,2020MNRAS.496.2155Z}, due to the different life cycles of different regions \citep[e.g.][]{2014MNRAS.439.3239K}, i.e.\ at least partially independent cloud assembly, collapse, star formation and disruption due to stellar feedback. A statistical analysis of CO and H$\alpha$ snapshots of these regions therefore allows us to estimate the characteristic lifetime of the molecular gas ($t_{\rm gas}$), the duration of the young stellar phase ($t_{\rm star}$), duration of the overlapping/feedback phase ($t_{\rm fb}$), as well as the characteristic separation length of individual regions ($\lambda$). To infer these quantities, we adopt here the methods developed by \cite{2018MNRAS.479.1866K} and use the associated {\tt HEISENBERG} code. This approach has been used previously to characterise the GMC life-cycle for a wide range of observed galaxies \citep[e.g.][]{2019Natur.569..519K,2020MNRAS.493.2872C,2020MNRAS.497.2286W,2020MNRAS.496.2155Z,2021MNRAS.504..487K,2022MNRAS.509..272C}, as well as for numerical simulations \citep[e.g.][]{2019MNRAS.487.1717F,2021MNRAS.505.3470J}.

We briefly summarise the methodology below, while the details can be found in Section~3 of \citet{2018MNRAS.479.1866K}. The first step is to measure CO-to-H$\alpha$ flux ratios in apertures of a range of sizes, all centred on the peaks of the CO and the H$\alpha$ emission, respectively. The identification of the molecular gas and ionised-gas peaks and the total flux calculations use the same maps as for the depletion time calculations (see Section~\ref{analysis}). The relative changes of the flux ratios compared to the galactic average, as a function of the aperture size, are governed by the three aforementioned timescales ($t_{\rm gas}$, $t_{\rm star}$ and $t_{\rm fb}$) as well as $\lambda$, as shown in \citet{2014MNRAS.439.3239K}. We fit the observed CO-to-H$\alpha$ flux ratio with the analytical model presented in \citet{2018MNRAS.479.1866K} to constrain $\lambda$ and the ratios $t_{\rm gas}/t_{\rm star}$ and $t_{\rm fb}/t_{\rm star}$. Assuming a reference timescale $t_{\rm star,~ref} = 4.3$~Myr for the isolated stellar phase (the appropriate timescale when using continuum-subtracted H$\alpha$ emission as the SFR tracer; see \citealt{haydon2020uncertainty}), we convert these timescale ratios into absolute values.

\begin{table}
  \centering
  \caption{Best-fitting timescales and separation length.}
  \label{tab:tun}
  \begin{tabular}{|l|c|c|c|}
    \hline
    & Case~A & Case~B & Case~C \\
    \hline
    $t_{\rm gas}$ (Myr) & $7.1^{+5.3}_{-2.1}$ & $10.5^{+4.2}_{-2.6}$ & $6.7^{+5.0}_{-1.4}$\\
    \hline
    $t_{\rm fb}$ (Myr) & $\leq2.9$ & $\leq3.3$ & $\leq1.5$\\
    \hline
    $t_{\rm star}$ (Myr) & $5.5^{+1.5}_{-0.9}$ & $5.9^{+1.7}_{-0.9}$ & $4.4^{+1.4}_{-0.16}$\\
    \hline
    $\lambda$ (pc) & $145^{+59}_{-38}$ & $148^{+53}_{-28}$ & $163^{+33}_{-21}$\\
    \hline
  \end{tabular}
  
  {\it Notes:} {all values are associated with their $1\sigma$ uncertainties}
\end{table}

\begin{figure*}
  \centering
  \includegraphics[width=0.65\textwidth]{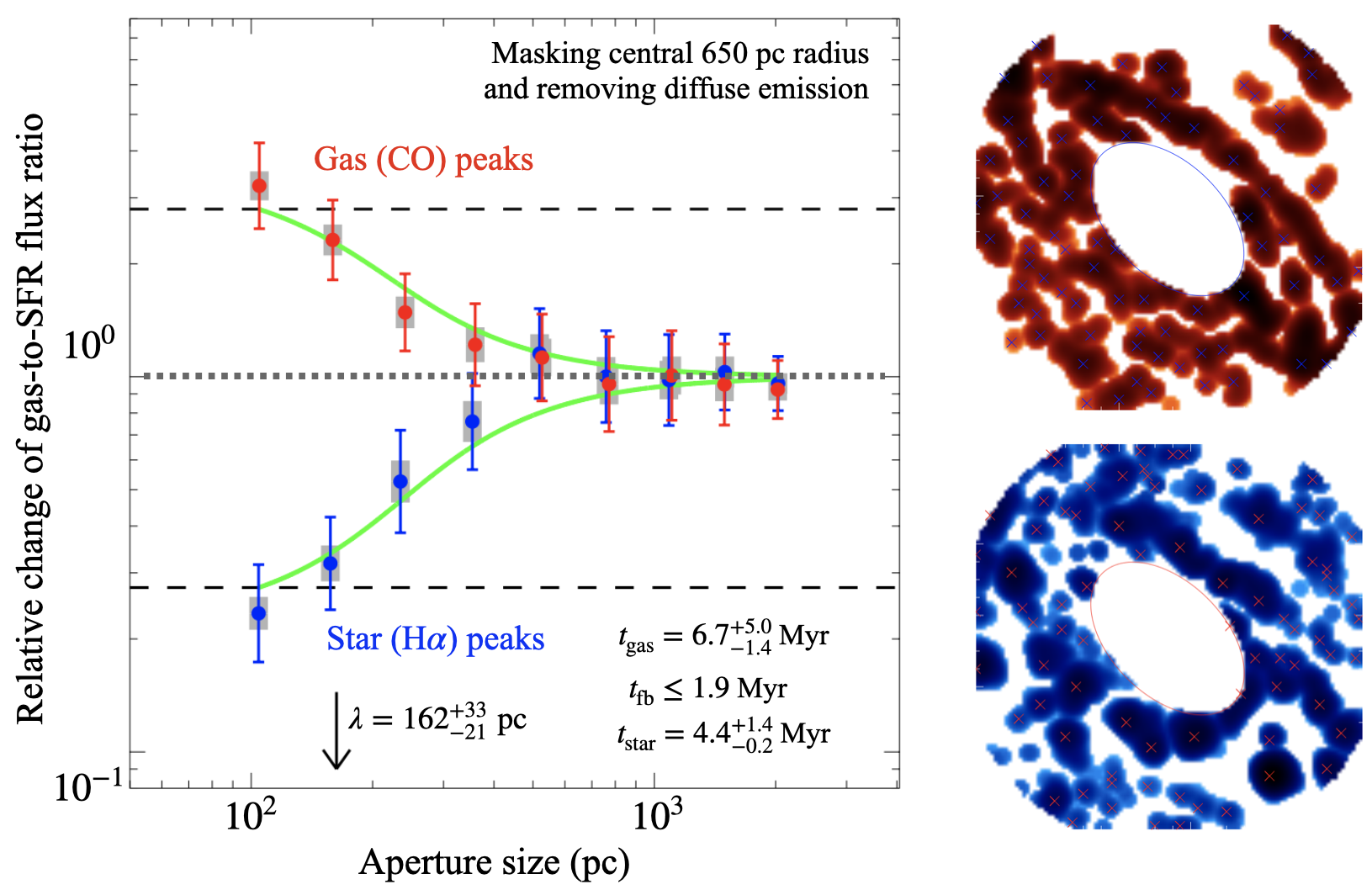}
  \caption{\label{fig:tuningfork} Left: relative change of the CO-to-H$\alpha$ flux ratio as a function of aperture size for Case C (see Section~\ref{dis:tuning}, with a focus on the molecular gas peaks (red datapoints) and the ionised-gas peaks (blue datapoints). The best-fitting model based on the uncertainty principle for star formation of  \citet{2018MNRAS.479.1866K} is shown as a green solid line (see Section~\ref{dis:tuning}). The horizontal grey dotted line indicates the galactic average of the flux ratio normalized to the mean $\tau_{\rm dep}=0.9$~Gyr within the ALMA FOV. The associated best-fitting parameters with $1\sigma$ uncertainties are listed at the bottom of the panel. Right: corresponding CO (top) and H$\alpha$ (bottom) maps.}
\end{figure*}

As the input maps for this analysis require contiguous emission across each regions, a naive separation of star-forming and AGN-ionised regions based on BPT diagram diagnostics would not work. Considering this constraint, we adopt a different approach to nevertheless attempt to distinguish emission arising from each of the three ionisation mechanisms discussed in Section~\ref{analysis} (i.e.\ star-forming, composite and AGN-ionised regions). For Case~A listed in Table~\ref{tab:tun}, all emission within the field of view is considered, implying that all three ionisation mechanisms are included. For Case~B, a central region of $650$~pc radius is masked out. As the majority of AGN-ionised regions is within this central region, Case~B includes mostly star-forming and composite regions. For Case~C, in addition to removing the central region, diffuse emissions is also removed across the field, achieved by tapering large-scale emission in Fourier space. More specifically, the emission in each line is transformed into Fourier space, and a Gaussian high-pass filter is then applied centred at zero Fourier frequency, using the best-fitting $\lambda$ (the separation lengthscale) in an iterative process \citep{hygate2019uncertainty}. As can be inferred from Figure~\ref{fig:BPT_full} and the discussions in Sections~\ref{line_ratio} and \ref{dis:AGN_scope}, diffuse emission is dominated by ionisation from the AGN, generally leading to a composite classification. Case~C thus mostly contains star-forming regions, as required by the adopted methodology, and therefore constitutes our reference case. The Case~C maps, relative flux ratios and best-fitting model from the {\tt HEISENBERG} code are shown in Figure~\ref{fig:tuningfork}, while those for Cases~A and B are shown in Figure~\ref{fig:tuningfork_app} for reference.

As listed in Table~ \ref{tab:tun}, the fitting results from the three cases agree with one another within the $1\sigma$ uncertainties. The fact that the separation lengths are $1.5$ times larger than the spatial resolution of our data (i.e.\ $\lambda>100$~pc) gives us confidence that the spatial resolution of our maps is sufficient for this analysis. 
The separation length of $\approx160$~pc is also typical of nearby spirals ($100$--$300$~pc; \citealt{2020MNRAS.493.2872C}). The molecular gas lifetime is rather short ($t_{\rm gas}=6.7^{+5.0}_{-1.4}$~Myr), but with high upper limit. A gas lifetime of $\approx10$~Myr, comparable to the representative cloud free-fall time, typically indicates that the cloud is regulated by internal dynamical processes. Comparing to the GMC lifetimes of other nearby galaxies \citep[e.g.][]{2019Natur.569..519K,2020MNRAS.496.2155Z,2022MNRAS.509..272C}, the cloud lifetime is reasonable, given the small galactocentric radii where the dynamical timescales are typically small. However, the number of flux peaks identified, 77 for CO flux and 80 for H$\alpha$ flux, is not large enough to constrain timescale radial trends. The blending of the CO and H$\alpha$ peaks also results in significant uncertainties on the feedback timescale $t_{\rm fb}$. This is characterized with a region filling factor $\zeta$ described in \citet{2018MNRAS.479.1866K}, which is $\approx0.7$ in our best-fitting result that prevent the accurate measurement of $t_{\rm fb}$ within a few Myr \citep[see e.g.][]{2022MNRAS.509..272C}. Therefore, only the upper limits of $t_{\rm fb}$ for each case are listed in Table~\ref{tab:tun}. These factors unfortunately prohibit firm conclusions about the nature of the dominant GMC-scale feedback and quenching mechanisms in the central region of NGC~3169. Observations with higher spatial resolution and larger FOV can help to resolve some of the issues.


\section{Discussion}
\label{discussion}


\subsection{Quenching from bulge dynamics}
\label{dis:other}

In Section~\ref{radialtrend}, we showed that the short depletion times in the star-forming ring gradually increase as we probe close to the centre of the galaxy, and identified it as a quench of star-formation. A common explanation for star-formation quenching in galaxy bulges is the bulge dynamics itself, which can enhance velocity dispersions in the central regions of the galaxy \citep[e.g.][]{martig2009morphological,martig2013atlas3d}. In particular, \citet{gensior2020heart} and \citet{gensior2021elephant} simulated the influence of a bulge on the radial variations of the SFE. A power-law relation normally exists between $\Sigma_{\rm SFR}$ and $\Sigma_{\rm H_2}$ in bulgeless galaxies (i.e.\ the usual ``spatially-resolved'' Kennicutt-Schmidt relation; \citealt{kennicutt2012star,bigiel2008star}). A similar power-law relation is reproduced by the simulations of bulgeless galaxies in \citet{gensior2020heart}, but with a different slope. However, adding bulges to the galaxy models results in $\Sigma_{\rm SFR}/\Sigma_{\rm H_2}$ ratios that are systematically lower at small galactocentric radii, and the decrease is more extreme in denser bulges. 

\begin{figure}
\centering
\includegraphics[width=0.49\textwidth]{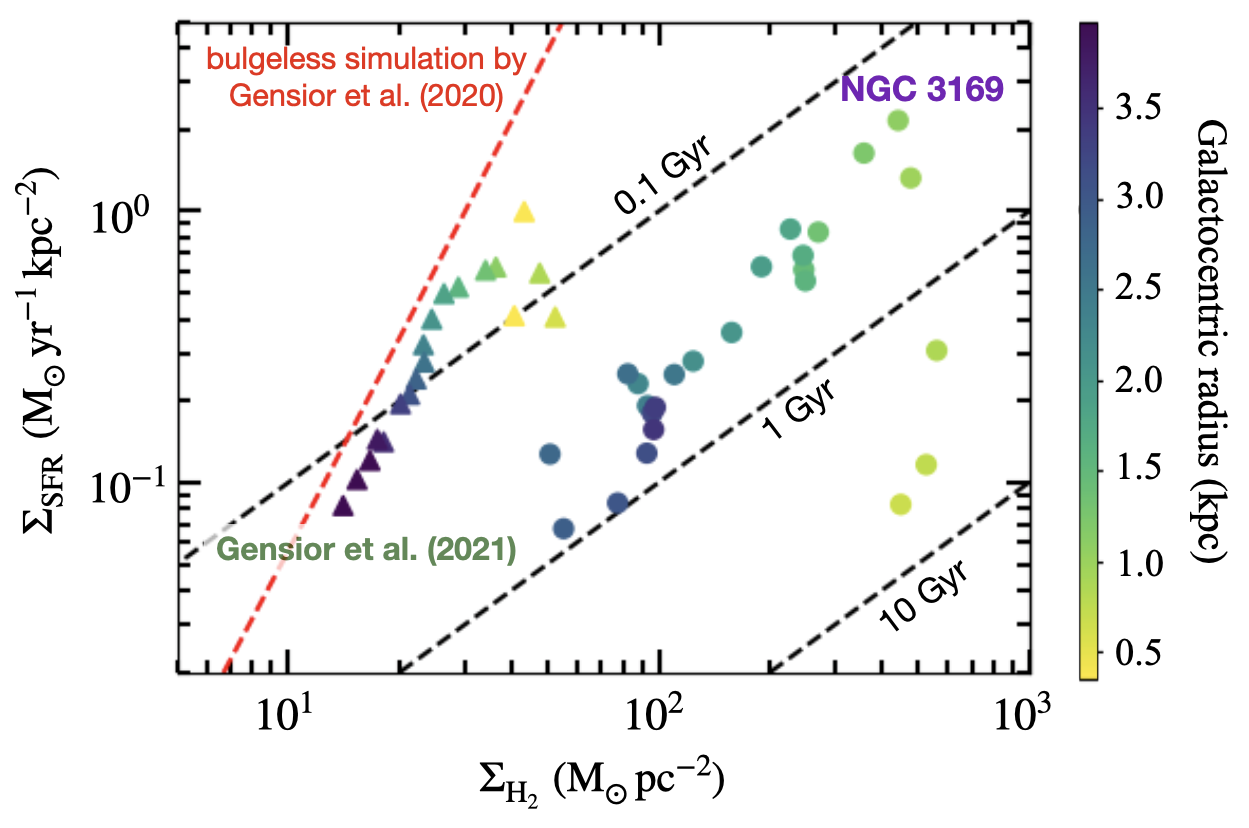}
\caption{\label{fig:other_bulge} Relation of $\Sigma_{\rm H_2}$-$\Sigma_{\rm SFR}$. The galactocentric radius of each datapoint is indicated by its colour scales. Filled circles are azimuthal averages from our NGC~3169 data, for spaxels classified as star-forming only. Filled triangles are azimuthal averages from the simulations of \citet{gensior2021elephant}. The red dashed line shows the power-law relation best-fitting the results from a bulgeless galaxy simulation \citep{gensior2020heart}. Representative depletion timescales of $0.1$, $1$ and $10$~Gyr are shown as black dashed lines.}
\end{figure}

Here, we attempt to compare the radial trend of $\Sigma_{\rm SFR}/\Sigma_{\rm H_2}$ of both our NGC~3169 data and simulations of galaxies with dense bulges. The $\Sigma_{\rm SFR}-\Sigma_{\rm H_2}$ ratio of NGC~3169 is shown in Figure~\ref{fig:other_bulge} as a function of galactocentric radius (using running means; another way to visualise Figs.~\ref{fig:dept_map}--\ref{fig:dept_rad}). The estimated power-law relation of bulgeless galaxies from the simulation of \citet{gensior2020heart} is shown as a red dashed line, and representative depletion timescales shown as black dashed lines. The bulge simulations of \citet{gensior2021elephant}, with an effective radius of $\approx1$~kpc and a bulge-to-total mass ratio of $\approx0.9$, are shown with triangles. As the simulations are idealised cases, with smooth rotation curves and a limited range of $\Sigma_{\rm H_2}$, they do not accurately represent the bulge environment of NGC~3169. Nevertheless, the simulations do show that the presence of a bulge can produce a radial trend of the $\Sigma_{\rm SFR}/\Sigma_{\rm H_2}$ ratio similar to that observed in NGC~3169. We also notice that the simulated bulges are much denser than the bulge of NGC~3169, with an effective radius $\approx1.4$~kpc, a bulge-to-total mass ratio $\approx0.1$ and B/D~$>1$ only within a radius of $660$~pc. This suggests that other mechanisms may be required to quench SF within the central kpc of this galaxy.

Using the relation found by \citet{gensior2021elephant}, the sSFR predicted from bulge dynamics can be inferred by relating it to the total molecular gas-to-stellar mass fraction ($f_{\rm gas}$) and the stellar mass surface density ($\mu_\star$):
\begin{equation}
\begin{split}
    \log({\rm sSFR_{sim}\,/\,Gyr^{-1}})=-1.79+4.00\log\left(\frac{f_{\rm gas} }{0.05}\right)\\
    ~~~~~~~~~~~~~~~~~~~~~~~~~~~~~~~~-0.99\log\left(\frac{\mu_\star}{10^8\,{\rm M_\odot\,kpc^{-2}}}\right)~.
\end{split}
\end{equation}
Using the total molecular gas mass of $\approx10^{9.3}~{\rm M_\odot}$, the total stellar mass of $\approx10^{10.8}~{\rm M_\odot}$ and the stellar surface density of $\approx10^{8.3}~{\rm M_\odot}$~kpc$^{-2}$ (all listed in Table~\ref{tab:summary}), we obtained a prediction of $\log({\rm sSFR_{sim}\,/\,Gyr^{-1}})=-0.68$. Comparing to the observed $\log({\rm sSFR[Gyr^{-1}]})=-1.55$, using the total SFR of $\approx10^{0.29}~{\rm M_\odot}$~yr$^{-1}$ (in Table~\ref{tab:summary}), this suggests that bulge dynamics are not the sole source of SF suppression. Other factors could be acting in concert with the dynamics to further decrease the sSFR. 


\subsection{AGN feedback}
\label{dis:AGN_scope}

In Section~\ref{line_ratio}, with the aid of BPT diagrams, we analysed the optical emission line ratios, across the whole disc of NGC~3169. In addition to using these to separate star-formation, composite and AGN ionisation, which help to interpret the (often large) inferred molecular gas depletion times, we established that AGN ionisation is pervasive up to a galactocentric radius of $\approx2$~kpc. It is therefore natural to investigate whether other lines of evidence support a causal relationship between the LLAGN and SFE.

One important observable is the coincidence of a large velocity dispersion (in both ionised and molecular gas) and AGN ionisation in many spaxels. Indeed, as shown in Figures~\ref{fig:Halpha_data} and \ref{fig:CO_maps}, the velocity dispersions of both H$\alpha$ and CO are significantly larger in the inner disc than in the star-forming ring (i.e.\ at radii~$\lesssim650$~pc), while Figure~\ref{fig:BPT_full} clearly shows that these same regions are AGN ionised, potentially indicating shock heating and/or ionisation from the AGN radiation. We can further probe this velocity dispersion-ionisation mechanism correlation with star-formation quenching by looking back at the \ion{H}{ii} region dynamics discussed in Section~\ref{virial}. There, we uncovered a weak correlation between the \ion{H}{ii} regions' virial parameters and the molecular gas depletion times, higher virial parameters being associated with longer depletion times (or equivalently lower SFEs). As the velocity dispersion is the dominant parameter in virial parameter calculations, there is in turn a correlation between these high velocity dispersions and suppressed star formation. However, it is important to note that beam smearing may have a large impact here due to the rapid changes of rotation curve slopes shown in Section~\ref{vfield}. 

As discussed in Section~\ref{CO} and \ref{vfield}, there are traces of AGN-driven outflows that can contribute to the enhanced velocity dispersions discussed above, quenching the star formation in the innermost regions of NGC~3169. Similar negative feedback mechanisms are discussed in recent works of \citet{venturi2021magnum} and \citet{gao2021nuclear}. On the other hand, we also identified a fraction of AGN-ionisation in the star-forming ring, associated with diffuse emissions surrounding the H$\alpha$ peaks (see Section~\ref{line_ratio}). It suggests positive feedback from the AGN that enhanced SF in the star-forming ring. Overall, AGN feedback could thus be partly responsible for the radial profile of SFE from the star-forming ring to the central $100$~pc in radius of NGC~3169. 


\subsection{Quenching from the bar}

As shown in Section~\ref{vfield}, there are multiple sources of non-circular motions in the central region of NGC~3169. An outstanding one is significant spiral-like structures that can be associated with inflow/outflow streaming motions as a result of the small-scale bar. Barred galaxies often have characteristic star-formation features, as studied in great details in the recent work of \citet{sato2021relating}. In particular, there are often cold gas overdensities and enhanced star formation both in a nuclear ring, caused by resonances with the bar, and at the nodes where the spiral arms and bar connect. With material tranported into the bulge by inflows within the bar, turbulence can be enhanced in the central region and star formation may be suppressed. However, simulations of clouds and star formation in galactic bars \citep[in e.g.][]{renaud2015environmental} have also shown that enhanced turbulence can, in some cases, induce strong local compression and lead to star bursts. These simulations also illustrated that strong shear along the elongated orbits in galactic bars may instead suppresses turbulence and the associated local compression, while also stabilizing against gravitational instabilities.

The presence of the small-scale bar may explain the radial trend of $\tau_{\rm dep}$ observed in NGC~3169, although there is no corresponding overdensity of GMCs at the nodes. We should also note that the velocity dispersions of both the ionised and the molecular gas in NGC~3169 are not particularly high. As illustrated in many works \citep[e.g.][]{sun2020molecular,deconto2022ionised}, in the central kpc in radius, AGN-host galaxies can have velocity dispersions over $100$~km~s$^{-1}$ higher than those of other galaxies with similar morphologies. 

A more detailed analysis of the spatially-resolved properties of GMCs in NGC~3169 could also provide more information about the influence of the bulge dynamics and shear. For example, as part of the WISDOM project, \citet{liu2021wisdom} analysed individual GMCs in NGC~4429 and revisited the conventional Virial theorem, including the effects of the galatic gravitational potential in addition to the usual cloud self-gravity. Applying such a model to the bulge region of NGC~3169 may help to better understand the dynamics of individual GMCs and \ion{H}{ii} regions, and to examine the impact of energy injection from the AGN. However, the nominal spatial resolution of the current CO observations ($\approx65$~pc) is not sufficient for this purpose. It may still be worthwhile to probe giant molecular associations (GMAs; i.e.\ collections of GMCs), but wider spatial coverage would be necessary to draw firm statistical conclusions.


\section{Conclusions}
\label{conclusion}

In this work, we study the star formation efficiency of the central region of NGC~3169 ($\approx3.6\times3.6$~kpc$^2$). The molecular gas is traced by \textsuperscript{12}CO(2-1) observations from ALMA, while the star formation rate is traced by extinction-corrected H$\alpha$ observations from SITELLE at CFHT. The surface densities of molecular and ionised gas are measured at a spatial resolution of $<100$~pc. The depletion times and $\Sigma_{\rm SFR}$-$\Sigma_{\rm H_2}$ relation are measured accordingly and probed for variations with galactocentric distance and bulge dynamics. To explain the enhanced star-formation suppression closer to the galactic centre, we examined the potential impact of AGN feedback and other mechanisms. Our key findings are summarised as follows:

\begin{enumerate}

\item We identified star-forming, AGN-ionised and composite regions based on emission-line ratios, and showed that the AGN impact is strongest within the inner disc, while composite regions reach over $2$~kpc from the galaxy centre. For the AGN-ionised and composite regions, we estimated upper limits of $\Sigma_{\rm SFR}$ and hence lower limits of depletion times. 

\item The overall star-formation relation follows closely the power-law relations of \citet{kennicutt1998global} and \citet{bigiel2008star}, with depletion times ranging from $0.1$ to $10$~Gyr. For star-forming regions, the peak of $\Sigma_{\rm SFR}-\Sigma_{\rm H_2}$ distribution is slightly above 2 Gyr, indicating star-formation quenching. For composite and AGN-ionised regions, the distributions have different peaks with much larger $\Sigma_{\rm SFR}$ and slightly larger $\Sigma_{\rm H_2}$, which could however result from $\Sigma_{\rm SFR}$ overestimates due to the unkown fraction of star-forming ionisation in these spaxels.

\item Ionised-gas emission appears to be concentrated within a star-forming ring, while molecular gas fills up both a smaller central disc and the ring. The depletion times in the star-forming ring are $\approx0.3$~Gyr, comparable to typical starbursts. As the radius decreases within a galactocentric distance of $\approx0.9$~kpc, star formation becomes scarce while molecular gas gets denser, resulting in a longer depletion time. In star-forming regions, $\tau_{\rm dep}$ increases gradually from $\approx0.3$~Gyr in the starburst ring to $\approx2.3$~Gyr at a radius of $500$~pc.

\item There is a weak correlation between the molecular gas depletion times and the virial parameters of the associated \ion{H}{ii} regions. As large virial parameters are generally associated with high velocity dispersions, this suggests that the SFEs in the bulge of NGC~3169 are affected by turbulence. Potential sources of turbulence in this region are outflows from the AGN and inflows from the spiral arms to the inner bar. The molecular gas lifetime and feedback durations are constrained by the uncertainty principle for star formation of \citet{2018MNRAS.479.1866K} are reasonable comparing to other nearby galaxies.
Bulge dynamics may also contribute to the star-formation quenching in the central region.

Spatially resolved analyses of GMCs and \ion{H}{ii} regions are underway to better understand the kinematics and local environments of the bulge of NGC~3169 and similar galaxies. We are also exploiting other SFR indicators such as free-free emission from VLA, to minimise any uncertainty caused by dust extinction. Most importantly, to piece together the puzzle of star formation quenching at the heart of galaxies, we require more observational data probing galaxy bulges and ETGs.

\end{enumerate}

\section*{Acknowledgements}

We thank the referee, Frederic Bournaud, for useful comments that improved the analyses and manuscript. We thank Federico Lelli for the help with {\tt 3DBarolo} and valuable discussions. This research is based on observations obtained with the SITELLE instrument on the Canada-France-Hawaii Telescope (CFHT) which is operated from the summit of Maunakea, and the Atacama Large Millimeter/submillimeter Array (ALMA) in the Atacama desert. We are grateful to the CFHT and ALMA scheduling, data processing and archive teams. We also wish to acknowledge that the summit of Maunakea is a significant cultural and historic site for the indigenous Hawaiian community, while the the high-altitude plateau Chajnantor on which the ALMA telescope sits is sacred to indigenous Likanantai people. We are most grateful to have the opportunity of observing there. This paper also makes use of observations made with the NASA/ESA Hubble Space Telescope and the NASA/IPAC Extragalactic Database (NED). 

AL, HB, DH and LD acknowledge funding from the NSERC Discovery Grant and the Canada Research Chairs (CRC) programme. HB is grateful for support from the Natural Sciences and Engineering Research Council of Canada (NSERC) Alexander Graham Bell Canada Graduate Scholarship. LC acknowledges support by the ANID/FONDECYT Regular Project 1210992. MC and JMDK gratefully acknowledge funding from the German Research Foundation (DFG) through an Emmy Noether Grant (grant number KR4801/1-1), as well as from the European Research Council (ERC) under the European Union's Horizon 2020 research and innovation programme via the ERC Starting Grant MUSTANG (grant agreement number 714907). TAD and IR acknowledge support from the UK Science and Technology Facilities Council through grants ST/S00033X/1 and ST/W000830/1. JG gratefully acknowledges financial support from the Swiss National Science Foundation (grant no CRSII5\_193826). TGW acknowledges funding from the European Research Council (ERC) under the European Union’s Horizon 2020 research and innovation programme (grant agreement No. 694343).  

\section*{Data Availability}
The raw data underlying this article are publicly available on NRAO and CFHT archive. All analysed data are available in the article and upon request.

\bibliographystyle{mnras}
\bibliography{ref_v2}

\newpage

\appendix

\section{Ionisation-mechanism classification in a three-dimensional line ratio space}
\label{app_3DBPT}

\begin{figure*}
  \centering
  \includegraphics[width=0.98\textwidth]{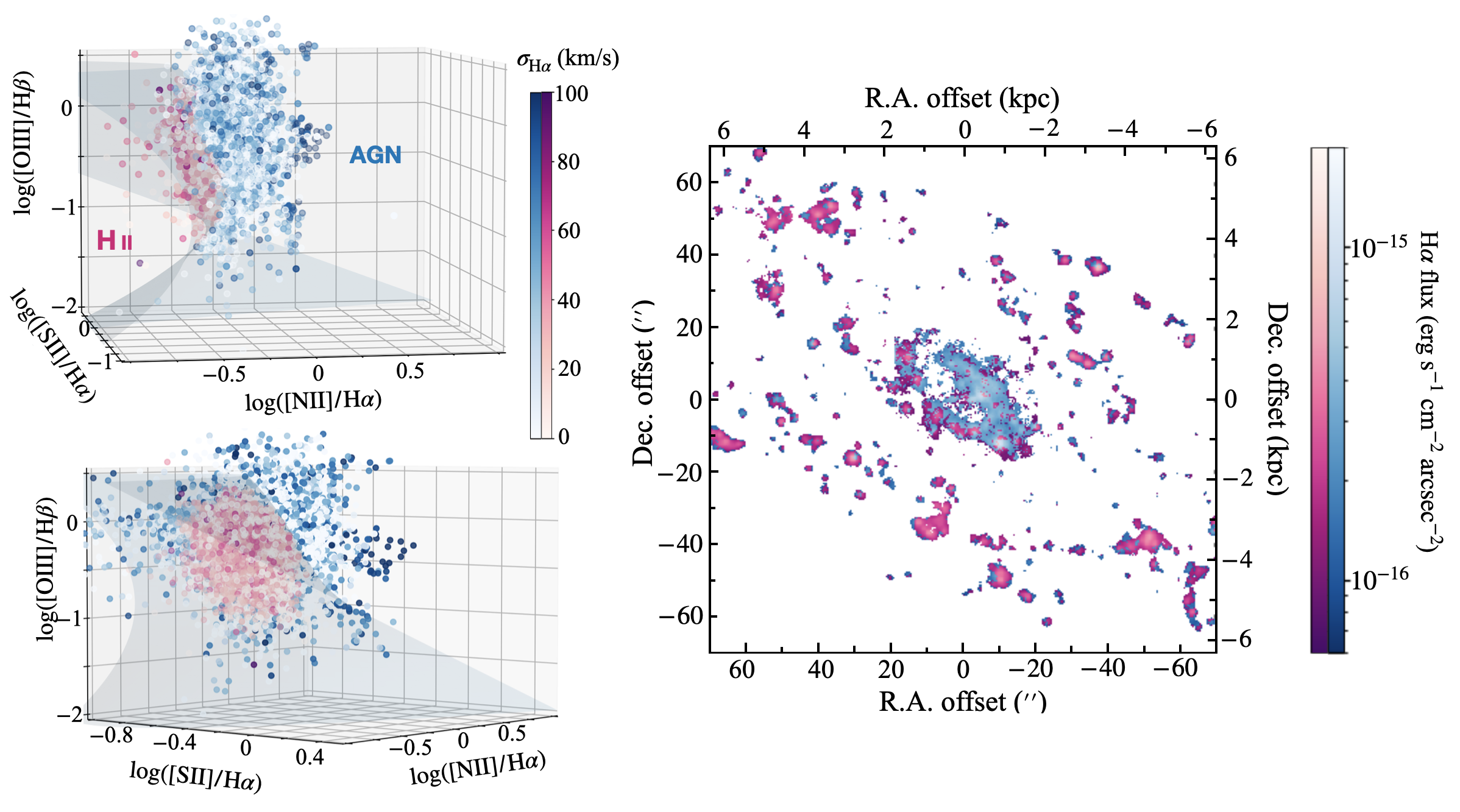}
  \caption{\label{fig:3DBPT} Ionisation-mechanism classification over the whole disc of NGC~3169 in a three-dimensional line ratio space. Left: three-dimensional plot of the line ratios [\ion{O}{iii}]/H$\beta$, [\ion{N}{ii}]/H$\alpha$ and [\ion{S}{ii}]/H$\alpha$. AGN/LINER-ionised (blue) and \ion{H}{ii} (pink) regions are separated using a plane defined by equation (11) and Table 2 in \citet{law2021sdss}. The color gradient shows the broadening of the H$\alpha$ line, indicative of gas temperature and associated ionisation mechanism. The top and bottom plots illustrate different directions of viewing the same parameter space. Right: the H$\alpha$ flux map colour-coded according to the region separation shown in the left panels, as a comparison to the right panels of Figure~\ref{fig:BPT_full}.}
\end{figure*}

\clearpage

\section{HII catalogue}
\label{app_HIIregion}

\begin{table*}
  \caption{Catalogue of all \ion{H}{ii} region parameters discussed in Section~\ref{virial}.}
  \label{tab:HII}
  \begin{tabular}{ccccccccccc}
    \hline
    ID & R.A. & Dec. & $R_{\rm \ion{H}{ii}}$ & $R_{\rm G}$ & $\log(L_{\rm \ion{H}{ii}} /$ergs s$^{-1}$) & $M_{\rm \ion{H}{ii}}$ & $V_{\rm avg,\ion{H}{ii}}$ & $\sigma_{\rm \ion{H}{ii}}$ & $\log(\alpha_{\rm vir,\ion{H}{ii}})$ & $\tau_{\rm dep,\ion{H}{ii}}$ \\
    -- & (deg) & (deg) & (pc) & (kpc) & -- & ($\times10^3$~M$_\odot$) & (km s$^{-1}$) & (km s$^{-1}$) & -- & (Gyr) \\
    \hline
    0 & 153.565 & 3.463 & <100 & 2.61 & 36.8 & 46 & 1360 & N/A & N/A & N/A \\
    1 & 153.569 & 3.463 & <100 & 2.71 & 36.8 & 46 & 1548 & 68 & 4.11 & N/A \\
    2 & 153.568 & 3.463 & <100 & 2.42 & 36.9 & 50 & 1492 & 30 & 3.36 & N/A \\
    3 & 153.562 & 3.464 & <100 & 3.26 & 37.0 & 55 & 1303 & N/A & N/A & 0.39 \\
    4 & 153.569 & 3.464 & 145 & 2.53 & 37.7 & 197 & 1497 & 32 & 2.94 & N/A \\
    5 & 153.568 & 3.464 & 237 & 2.07 & 37.6 & 349 & 1495 & 39 & 3.08 & 0.69 \\
    6 & 153.567 & 3.464 & 193 & 1.79 & 38.6 & 845 & 1470 & 28 & 2.33 & 0.19 \\
    7 & 153.568 & 3.465 & 146 & 2.08 & 38.3 & 373 & 1498 & 21 & 2.29 & 0.45 \\
    8 & 153.564 & 3.465 & 187 & 1.73 & 38.2 & 489 & 1330 & 21 & 2.31 & 0.16 \\
    9 & 153.562 & 3.465 & 138 & 2.42 & 38.1 & 272 & 1278 & 21 & 2.42 & 0.14 \\
    10 & 153.565 & 3.465 & 249 & 1.34 & 38.3 & 861 & 1397 & 33 & 2.57 & 0.12 \\
    11 & 153.567 & 3.465 & 154 & 1.43 & 37.7 & 206 & 1493 & 33 & 2.96 & 1.85 \\
    12 & 153.569 & 3.465 & 138 & 2.07 & 38.0 & 270 & 1491 & 37 & 2.91 & 0.20 \\
    13 & 153.563 & 3.466 & 204 & 1.59 & 38.0 & 452 & 1303 & 19 & 2.30 & 0.14 \\
    14 & 153.565 & 3.466 & 205 & 1.15 & 38.3 & 658 & 1393 & 28 & 2.46 & 0.15 \\
    15 & 153.566 & 3.466 & 227 & 1.13 & 37.6 & 339 & 1474 & 52 & 3.32 & 1.67 \\
    16 & 153.570 & 3.466 & 147 & 2.56 & 38.2 & 351 & 1444 & 27 & 2.55 & N/A \\
    17 & 153.564 & 3.466 & 210 & 1.30 & 37.9 & 421 & 1350 & 19 & 2.31 & 0.26 \\
    18 & 153.568 & 3.466 & 104 & 1.78 & 38.8 & 406 & 1466 & 33 & 2.52 & 0.28 \\
    19 & 153.561 & 3.466 & 167 & 2.26 & 37.9 & 305 & 1259 & 24 & 2.57 & 0.20 \\
    20 & 153.562 & 3.467 & 309 & 1.77 & 37.9 & 722 & 1251 & 24 & 2.45 & 0.15 \\
    21 & 153.560 & 3.466 & <100 & 2.62 & 36.8 & 46 & 1230 & N/A & N/A & 0.56 \\
    22 & 153.567 & 3.467 & 226 & 1.42 & 37.5 & 298 & 1423 & 29 & 2.88 & 0.40 \\
    23 & 153.567 & 3.467 & 179 & 1.00 & 37.7 & 273 & 1410 & 38 & 3.05 & 1.12 \\
    24 & 153.563 & 3.467 & 272 & 1.25 & 37.5 & 389 & 1284 & 22 & 2.61 & 0.70 \\
    25 & 153.568 & 3.467 & 186 & 1.63 & 37.7 & 289 & 1392 & 24 & 2.65 & 0.53 \\
    26 & 153.560 & 3.467 & 178 & 2.81 & 38.1 & 404 & 1226 & 16 & 2.10 & 0.21 \\
    27 & 153.561 & 3.467 & <100 & 1.83 & 37.2 & 74 & 1208 & 23 & 2.96 & 0.32 \\
    28 & 153.567 & 3.467 & <100 & 1.45 & 37.2 & 70 & 1379 & 37 & 3.39 & 1.29 \\
    29 & 153.562 & 3.467 & 335 & 1.53 & 37.4 & 491 & 1199 & 22 & 2.57 & 0.30 \\
    30 & 153.567 & 3.467 & <100 & 1.08 & 36.9 & 51 & 1389 & 17 & 2.88 & 1.06 \\
    31 & 153.560 & 3.468 & <100 & 2.18 & 36.8 & 46 & 1175 & N/A & N/A & 1.58 \\
    32 & 153.563 & 3.468 & <100 & 0.67 & 36.8 & 47 & 1190 & 90 & 4.34 & 0.79 \\
    33 & 153.564 & 3.468 & <100 & 0.11 & 36.9 & 50 & 1356 & 108 & 4.47 & 11.77 \\
    34 & 153.567 & 3.468 & <100 & 1.52 & 37.1 & 62 & 1348 & 39 & 3.49 & 0.94 \\
    35 & 153.566 & 3.468 & 181 & 0.81 & 38.7 & 808 & 1322 & 45 & 2.73 & 0.48 \\
    36 & 153.564 & 3.468 & <100 & 0.25 & 36.8 & 46 & 1158 & N/A & N/A & 5.40 \\
    37 & 153.567 & 3.469 & 200 & 1.56 & 38.4 & 740 & 1308 & 32 & 2.49 & 0.15 \\
    38 & 153.561 & 3.469 & <100 & 1.53 & 37.3 & 81 & 1130 & 26 & 3.02 & 1.23 \\
    39 & 153.561 & 3.469 & 227 & 1.86 & 38.2 & 657 & <1108 & 27 & 2.46 & 0.15 \\
    40 & 153.566 & 3.469 & 226 & 1.52 & 38.4 & 852 & 1278 & 29 & 2.40 & 0.14 \\
    41 & 153.565 & 3.470 & 213 & 1.24 & 37.7 & 360 & 1238 & 33 & 2.87 & 0.15 \\
    42 & 153.560 & 3.470 & 337 & 2.28 & 37.5 & 557 & 1123 & 39 & 3.02 & 0.09 \\
    43 & 153.565 & 3.470 & 281 & 1.37 & 37.5 & 440 & 1206 & 38 & 3.03 & 0.26 \\
    44 & 153.560 & 3.470 & 321 & 2.33 & 37.6 & 588 & 1079 & 29 & 2.73 & 0.01 \\
    45 & 153.569 & 3.470 & <100 & 3.55 & 36.8 & 46 & 1298 & N/A & N/A & N/A \\
    46 & 153.567 & 3.470 & 182 & 2.47 & 37.6 & 244 & 1259 & 28 & 2.85 & 0.24 \\
    47 & 153.563 & 3.471 & 342 & 1.42 & 37.4 & 470 & 1149 & 24 & 2.69 & 0.68 \\
    48 & 153.564 & 3.471 & <100 & 1.79 & 36.9 & 49 & 1195 & 33 & 3.46 & 0.56 \\
    49 & 153.560 & 3.471 & 198 & 2.29 & 38.6 & 829 & 1050 & 30 & 2.40 & 0.15 \\
    50 & 153.566 & 3.471 & <100 & 2.49 & 37.3 & 84 & 1251 & N/A & N/A & 0.02 \\
    51 & 153.559 & 3.472 & <100 & 2.66 & 37.8 & 146 & 1042 & 27 & 2.81 & N/A \\
    52 & 153.570 & 3.472 & <100 & 4.36 & 37.0 & 58 & 1303 & N/A & N/A & N/A \\
    53 & 153.563 & 3.472 & <100 & 2.06 & 36.8 & 47 & 1145 & 32 & 3.45 & 0.52 \\
    54 & 153.560 & 3.472 & <100 & 2.38 & 37.4 & 89 & 1046 & 14 & 2.43 & N/A \\
    55 & 153.570 & 3.472 & <100 & 4.80 & 36.8 & 46 & 1290 & N/A & N/A & N/A \\
    56 & 153.561 & 3.473 & <100 & 2.77 & 37.0 & 56 & 1088 & 23 & 3.09 & N/A \\
    57 & 153.562 & 3.473 & <100 & 2.74 & 36.8 & 46 & 1112 & 12 & 2.58 & N/A \\
  \end{tabular}

  {\it Notes:} {Column 2 and 3: R.A. and Dec. sky coordinates. Column 4: galactocentric radius ($R_{\rm G}$). Column 5: H$\alpha$ luminosity of the \ion{H}{ii} regions ($\log(L_{\rm \ion{H}{ii}})$). Column 6: ionised gas mass ($M_{\rm \ion{H}{ii}}$). Column 7: Mean line of sight velocity ($V_{\rm avg,\ion{H}{ii}}$). Column 8: velocity dispersion ($\sigma_{\rm \ion{H}{ii}}$). Column 9: virial parameter ($\log(\alpha_{\rm vir,\ion{H}{ii}})$). Column 10: depletion time $\tau_{\rm dep,\ion{H}{ii}}$. More parameters and/or parameters for the whole field of view of SITELLE can be provided upon request.}

\end{table*}

\begin{figure*}
  \centering
  \includegraphics[width=0.65\textwidth]{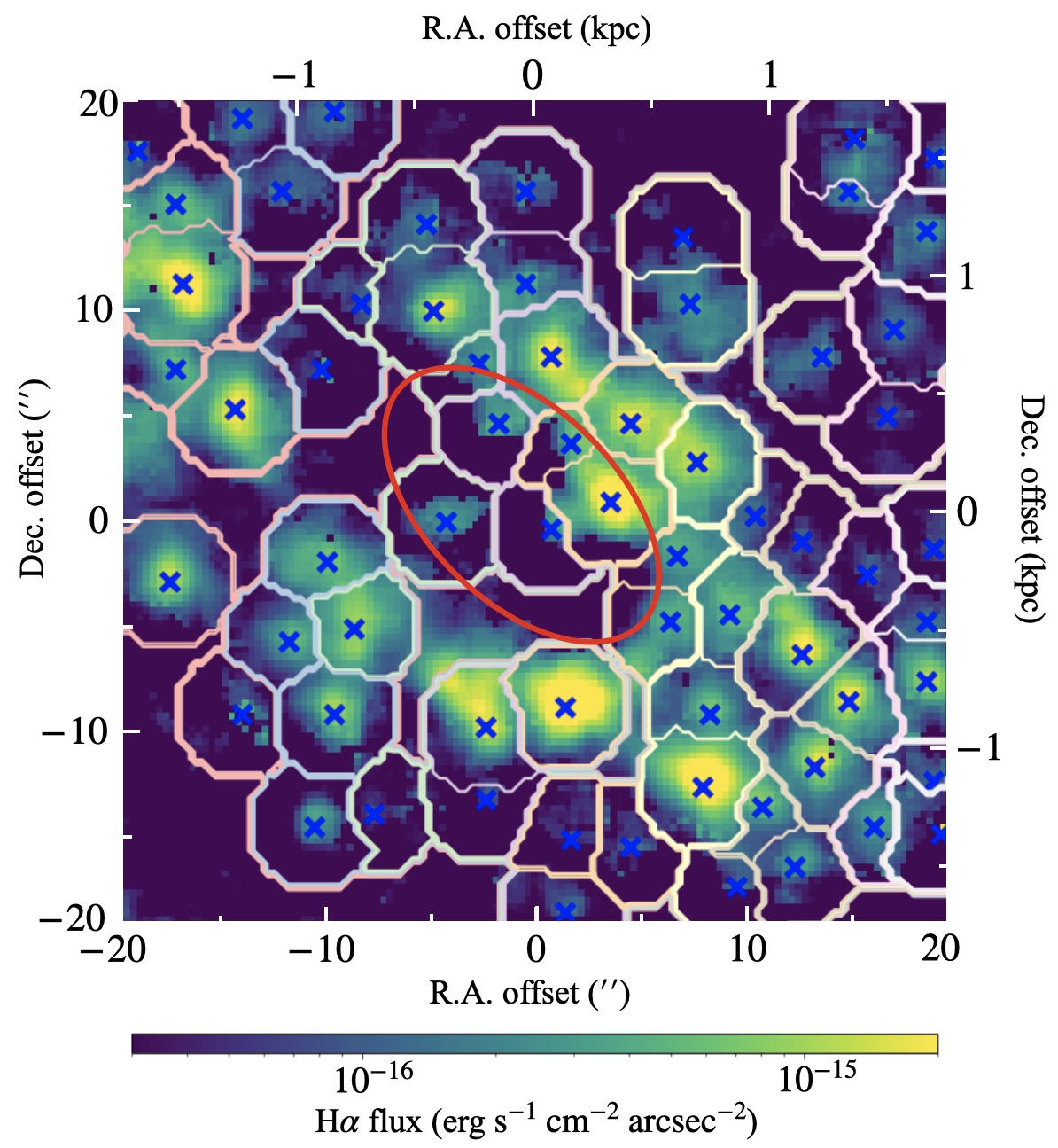}
  \caption{\label{fig:HII_app}Boundaries (white lines) and centres (blue crosses) of the \ion{H}{ii} regions identified in Section~\ref{virial}, overlaid on the H$\alpha$ flux map. The red ellipse separates the inner disc and outer star-forming ring discussed in Sections~\ref{analysis}--\ref{discussion}.}
\end{figure*}

\clearpage

\section{The best-fitting molecular gas kinematics model}
\label{app_CO_Vrad}

\begin{figure*}
  \centering
  \includegraphics[width=0.8\textwidth]{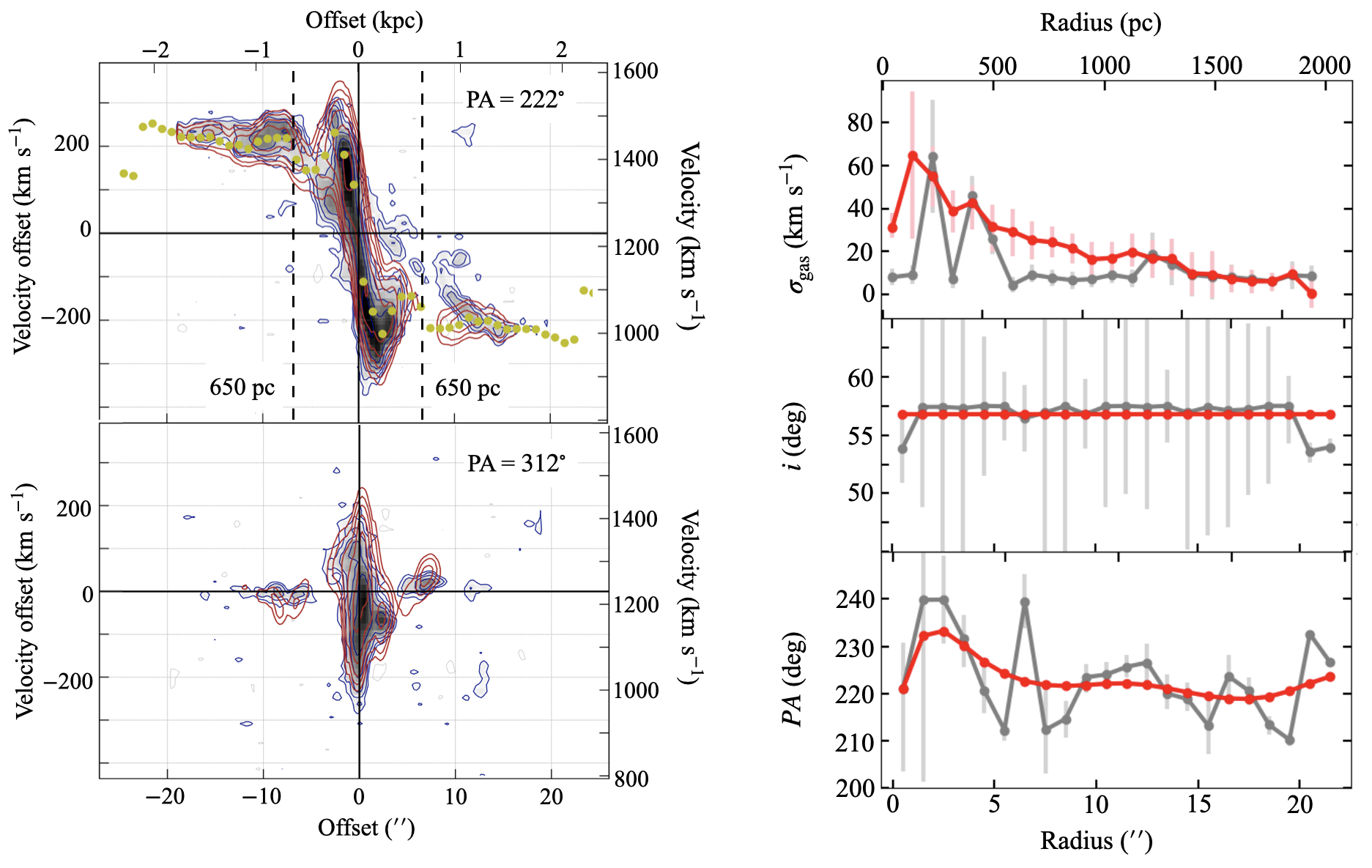}
  \caption{\label{fig:vel_field_app}Left: position–velocity diagrams of the CO emission along the major axis (top; at PA=$224^{\circ}$) and minor axis (bottom; at $314^{\circ}$). Shaded blue contours show the observed velocities, while red contours show the best-fitting models from {\tt 3DBarolo}. Overlaid yellow circles indicate the best-fitting rotation velocities. Right: radial profiles of the velocity dispersion ($\sigma_{\rm gas}$ (top), inclination ($i$; middle-right) and position angle ($PA$; bottom-right), from the best-fitting {\tt 3DBarolo} model assuming a fixed systemic (heliocentric) velocity of $1230$~km~s$^{-1}$. Grey datapoints are from a fit with all these parameters free, while red datapoints are from a fit with $i$ and $PA$ kept fixed.}
\end{figure*}

\clearpage

\section{Lifecycle of molecular clouds}
\label{app_tuning}

\begin{figure*}
  \centering
  \includegraphics[width=0.65\textwidth]{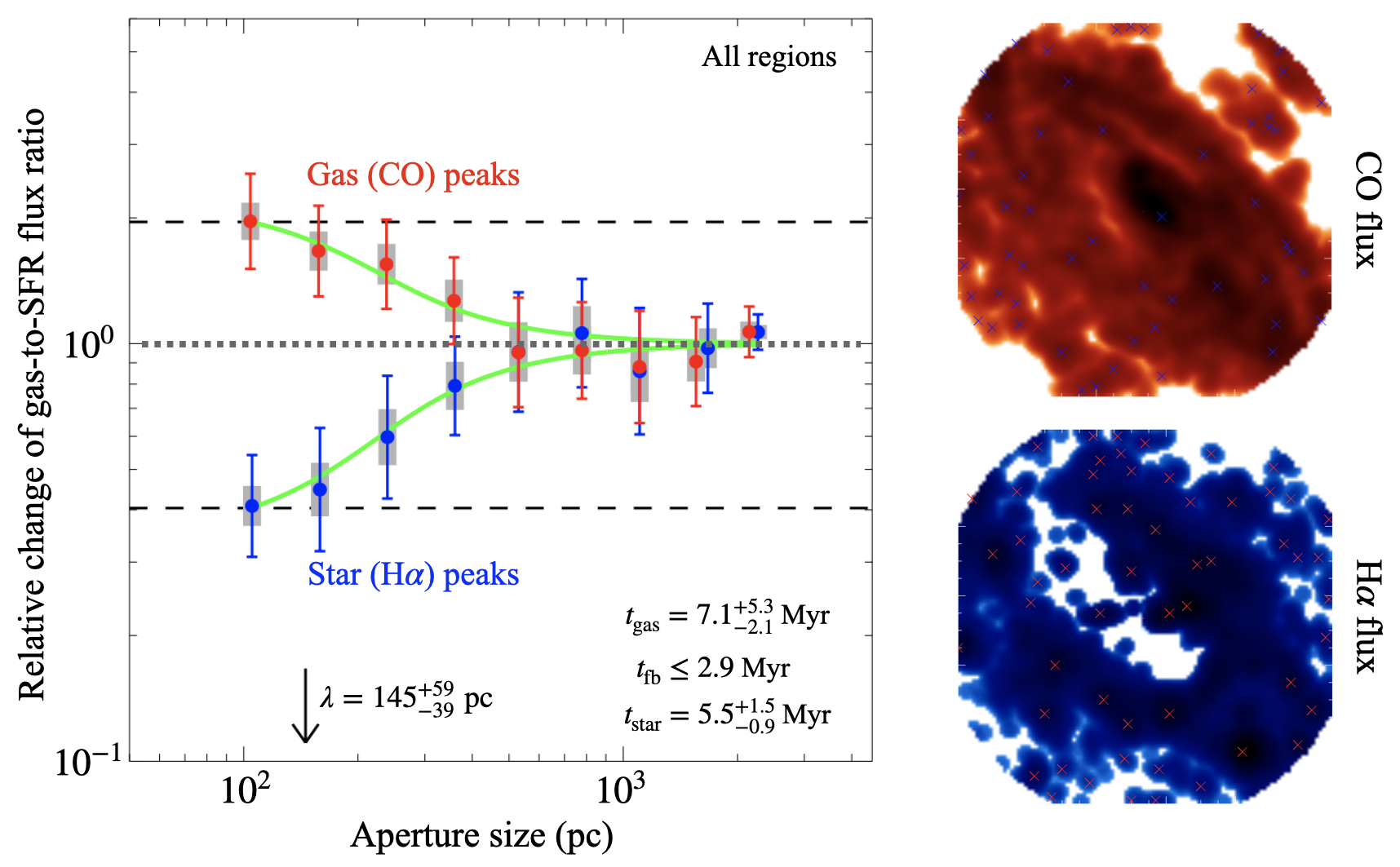}\\
  \vspace*{10mm}
  \includegraphics[width=0.65\textwidth]{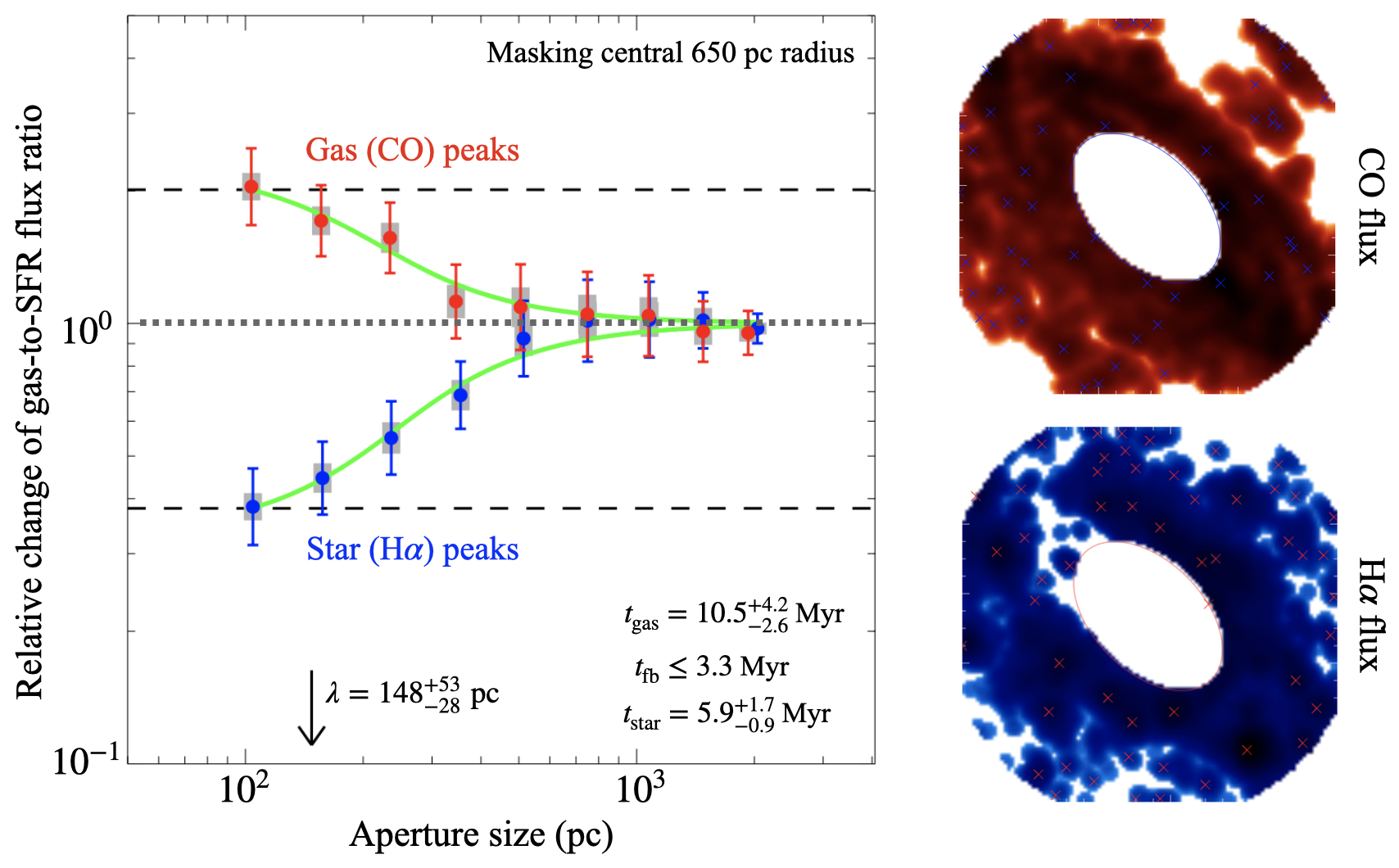}
  \caption{\label{fig:tuningfork_app}Same as Figure~\ref{fig:tuningfork} but for Cases~A (top) and B (bottom).}
\end{figure*}

\end{document}